\documentclass[nologo,11pt,a4paper]{ETHpaper}
\usepackage{graphicx, epsfig, amsmath, amssymb,color,wasysym}
\usepackage[sort&compress]{natbib}
\usepackage{caption}
\usepackage{subcaption}
\usepackage{longtable}
\usepackage{booktabs}
\usepackage{mathtools}
\usepackage{cancel}
\DeclarePairedDelimiter{\ceil}{\lceil}{\rceil}

\begin{document}

\newcommand{\mean}[1]{\left\langle #1 \right\rangle} \newcommand{\abs}[1]{\left| #1 \right|} \newcommand{\ul}[1]{\underline{#1}}
\renewcommand{\epsilon}{\varepsilon} \newcommand{\eps}{\varepsilon} \renewcommand*{\=}{{\kern0.1em=\kern0.1em}}
\renewcommand*{\-}{{\kern0.1em-\kern0.1em}} \newcommand*{\+}{{\kern0.1em+\kern0.1em}}

\title{Modeling User Reputation in Online Social Networks: \\
The Role of Costs, Benefits, and Reciprocity}

\titlealternative{Modeling User Reputation in Online Social Networks: The Role of Costs, Benefits, and Reciprocity}

\author{Frank Schweitzer$^{1,2,*}$, Pavlin Mavrodiev$^{1}$, Adrian M. Seufert$^{1}$,  David Garcia$^{2,3}$}

\authoralternative{F. Schweitzer, P. Mavrodiev, A. M. Seufert,  D. Garcia}

\address{$^1$ Chair of Systems Design, ETH Zurich, Weinbergstrasse 58, 8092 Zurich, Switzerland\\
$^2$Complexity Science Hub Vienna, Josefst{\"a}dter Strasse 39, 1080 Vienna, Austria\\
$^3$Section for Science of Complex Systems, CeMSIIS, \\ Medical University of Vienna, Spitalgasse 23, 1090 Vienna, Austria\\
$^*$ Corresponding author: \url{fschweitzer@ethz.ch}
}

\reference{(Submitted for publication)} 

\www{\url{http://www.sg.ethz.ch}}

\makeframing
\maketitle

\begin{abstract}

  We analyze an agent-based model to estimate how the costs and benefits of users in an online social network (OSN) impact the robustness of the OSN. 
  Benefits are measured in terms of relative reputation that users receives from their followers.
  They can be increased by direct and indirect reciprocity in following each other, which leads to a core-periphery structure of the OSN.  
  Costs relate to the effort to login, to maintain the profile, etc. and are assumed as constant for all users.
  The robustness of the OSN depends on the entry and exit of users over time. 
Intuitively, one would expect that higher costs lead to more users leaving and hence to a less robust OSN.
We demonstrate that an optimal cost level exists, which maximizes both the performance of the OSN, measured by means of the long-term average benefit of its users, and the robustness of the OSN, measured by means of the life-time of the core of the OSN. 
Our mathematical and computational analyses unfold how changes in the cost level impact reciprocity and subsequently the core-periphery structure of the OSN, to explain the  optimal cost level.

  \end{abstract}
\date{\today}

\section{Introduction}
\label{sec:Introduction}

Online social networks (OSN), like other types of social
organizations, undergo a steady evolution. 
New users enter the
network, while other users may decide to leave. 
New links between
users are formed online, based on shared information, friendship,
common interests, etc., and existing links may be deleted if users
leave, or commonalities have changed. 
Under normal circumstances, such events may not jeopardise the existence of the OSN, in
particular if the OSN is still popular and growing. 
However,
loosing users and links can pose a serious risk to OSN, even for large and successful ones on the scale of \texttt{Twitter} and \texttt{Facebook}. 
  As a recent empirical
  study  \citep{Garcia2013b} has shown that the
  dropout of some users can trigger cascades of other users leaving, which quickly accumulates to a level that threatens the existence
  of the OSN.

In order to investigate the emergence and the impact of such cascades
analytically and numerically, we propose a model that explicitly considers the costs and benefits of users of an OSN. 
\emph{Benefits} are
measured in terms of relative \emph{reputation} (see Sect. \ref{sec:costs}),
i.e. we take into account (i) the \emph{number} and the
\emph{direction} of links between users, and (ii) the impact of the
reputation of the counterparts on the reputation of a given user.
A gain of reputation can be seen as a benefit of being part of a social network. 
But there are also \emph{costs}
involved in being a member of an OSN, even if they are
implicit. These include, among others, the efforts to login, 
to maintain the individual profile, to learn how to efficiently use the GUI of the provider, or the effort to adapt
to changes in the GUI.

Our model extends previous notions of reputation in social networks which are based on the degree or the centrality of a user, measured by  traditional centrality metrics \citep{GayoAvello2013}.
Different from such topological metrics, we define the reputation of a user based on the reputation of those users that are linked to it. 
That means there is a \emph{value} assigned to incoming and outgoing links, as they increase either the own reputation or that of others.   
This implies  to consider (i) a \emph{directed network} and (ii) a self-consistent \emph{dynamics} of how reputation changes with the reputation of others. 

The assumption that individual reputation increases with the reputation of connected users is quite common in different ranking schemes. 
For example, the earliest version of \texttt{Google's} \texttt{PageRank} algorithm calculates the rank of a website as a function of the ranks of the sites linking to it.
Similarly, the vulnerability of financial institutions can be calculated dependent on the vulnerability of the conterparties connected to it \citep{battiston2012debtrank}. 
In more general terms, such an assumption follows (hetero)catalytic models
e.g. from chemistry and biology, where the concentration of a
particular (chemical) species depends on the concentration of those
species that produce, or feed, it \citep{Eigen.Schuster1979Hypercycleprincipleof}. 
I.e. models of prebiotic evolution or foodweb interactions have already utilized such assumptions. 

\citet{Jain1998} have combined  this dynamics with a network dynamics that
runs on a different time scale. 
At each time step (measured in network
time), the system was perturbed by an extremal dynamics where the least
performing node in the network was replaced by a new node that
randomly rewires itself back to the system. 
This model was already analyzed in detail \citep{Jain.Krishna2002CrashesRecoveriesand,SEUFERT2007} and extended to cover other phenomena,
such as strategic link formation between economic agents \citep{Koenig2009a}.

In this paper, we  build on the existing model class, but extend it in
a number of important points: (i) Different from the simple extremal
dynamics (where only one user is removed), we introduce a condition
for users to decide to leave, i.e. we allow \emph{many} more users to
leave dependent on their \emph{personal} cost-benefit ratio.
(ii) We focus on the effects that cascades of users leaving
  have on the robustness and the performance of the OSN.
\emph{Performance} is quantified using the long-term average reputation of users (see also Sect. \ref{sec:perform}) and is taken as a systemic measure for the OSN rather than as an individual one. 
\emph{Robustness} implies that the decision of users to leave the OSN does not trigger large cascades of other users leaving. 
We will proxy this by means of the life time of the core of the OSN (see also Sect. \ref{sec:network}). 

Our aim is not to focus on the \emph{size} of cascades, which was done in different network approaches to \emph{systemic risk} \citep{Lorenz2009,Schweitzer2013}, but rather to understand the impact that leaving users have on the OSN. 
As mentioned users will leave if their costs exceed their benefits. 
Different  from previous publications we \emph{vary} this cost, which is the  crucial parameter in our model. 
Interestingly, we find that both the robustness and the performance of the OSN are maximized for a \emph{non-zero cost} of usage.  

To explain this, we have to investigate how directed links impact the reputation across the OSN. 
This leads us to the problem of \emph{reciprocity}: If a user contributes to the reputation of \emph{others}, she may expect that these users in return also increase her own reputation. 
Mutual directed links  between two users, e.g. $1\to 2 \to 1$,  would indicate \emph{direct reciprocity}.
But users can also indirectly increase their reputation if they are part of a \emph{cycle}, e.g. $1\to 2 \to 3 \to 1$, which indicates \emph{indirect reciprocity}. 

While direct and indirect reciprocity is advantageous for the OSN from a user perspective, it is not well understood how reciprocity adds to systemic properties such as the robustness of the OSN. 
We will address this question by a mathematical analysis that explains the impact such cycles on the  core-periphery structure of the OSN.  
Users that are part of a cycle are well integrated in the OSN, maintain a high reputation and have little incentives to leave. 
But if users leave because of bad cost-benefit ratio, this will impact the number and sizes of such cycles, which we need to understand by means of an analytic approach that is complemented by computer simulations.

Our paper is organized as follows: 
Section \ref{sec:problem} introduces the model and the analytical framework to analyse the structure of an OSN. 
Different Appendices allow to study this framework by following concrete step-by-step examples.
In   Section \ref{sec:simulat} we present the
  results of computer  simulations, to demonstrate the emergence of an optimal cost
value, and we discuss the impact of different model parameters on the
structure and the performance of the OSN.
A discussion of the general insights and concluding remarks are provided in Section
  \ref{sec:disc}.  

\section{A reputation model}
\label{sec:problem}

\subsection{Costs and benefits}
\label{sec:costs}

Why do users leave a online social network (OSN)? A rational answer should be, they leave because their costs of staying in the
network exceed,  at a given point in time, their benefits of being members. This can be expressed by the
dynamics:
\begin{equation}
  \label{eq:stay}
  s_{i}(t)=\Theta\left[b_{i}(t)-c_{i}(t)\right]
\end{equation}
Here, $s_{i}(t)$ characterizes the current state of user $i$ at time $t$ as a binary variable: $s_{i}=1$ means that the user is
part of the network and $s_{i}=0$ means that at time $t$ the user leaves. $\Theta[z]$ is the Heavyside function which returns 1 if
$z \ge 0$ and 0 if $z < 0$. Thus, the current state of user $i$ depends on the difference between its benefits $b_{i}(t)$ and costs
$c_{i}(t)$.

In the following, we assume that the benefits of a user to join, and to stay, in the OSN result from the \emph{reputation}
$X_{i}(t)$ that the user receives from being connected to other users in the OSN. There are many social networks that operate this
way: in \texttt{Twitter} users get a reputation from the number of their followers, in product review communities like
\texttt{Amazon} or \texttt{Youtube} users earn their reputation from the votes of other users. 
In our model, we
particularly assume that the reputation of a user, $i$, does not just depend on the sheer number of other users that follow $i$, but
also on their reputation. I.e. if a user $j$ with a high reputation $x_{j}(t)$ connects to user $i$, the latter receives more
reputation than from a follower with low reputation. 

Such a reputation measure can be explicitly displayed on the site, like the
\texttt{Reddit} karma or the \emph{RG score} of \texttt{Researchgate}, which
increases with the reputation of followers and the feedback of the community.
On the other hand, user reputation can be implicit and not part of a user
profile, but can still be perceived through the activity of other users.
Examples of this implicit reputation are \emph{retweets} in \texttt{Twitter}
and \emph{likes} in \texttt{Facebook}.

We can express the impact of the followers on the reputation of a user by the following dynamics:
\begin{equation}
  \frac{dX_i}{dt}=\sum_{j=1}^{N} a_{ij}(t) X_j(t) -\phi X_i(t) 
  \label{eq:dyn}
\end{equation}    
The coefficients $a_{ij}$'s are elements of the adjacency matrix of
the OSN, \textbf{A}. They represent the link between users $j$ and $i$
in the OSN at time $t$. These are unweighted, but directed links,
because it makes a difference whether user $j$ follows user $i$, or the
other way round.  $a_{ij}(t)=1$ if there is a link from $j$ to $i$,
i.e. $j$ is called a \emph{follower} of $i$, and $a_{ij}(t)=0$
otherwise. Since a user cannot follow herself, we
set $a_{ii}(t) = 0$ for all $t$. 

The sum in Eq. (\ref{eq:dyn}) is over
all users that can potentially link to $i$. 
That means if a number of users leave the
OSN, they are replaced by the same number of new users joining the
OSN. 
This way  the total number of users, $N$, is kept constant. 
This is a first, and not necessarily the most realistic,
approximation to consider an \emph{entry and exit} dynamics, which can
be refined in subsequent investigations.  The second term on the
r.h.s. of Eq. (\ref{eq:dyn}) takes into account the effort to keep a
certain level of reputation. Without a number of followers, the
reputation of user $i$ cannot be maintained and thus decays
exponentially over time with a rate $\phi$. The larger $\phi$, the
higher the effort to maintain the reputation.

In order to relate reputation to the benefits of users in the OSN, we could simply assume that benefits grow in proportion to
reputation. On the other hand, the costs to stay in the OSN can be assumed to be a constant $\tau$ equal for all users, which can
be seen as the effort to login and stay active. 
Note that there is no cost involved in maintaining links.  According to Eq. (\ref{eq:dyn}) and
dependent on the specific social network, the reputation of users can grow to large numbers, while on the other hand it can become
infinitely small, but never zero.  Because we are more interested in the reputation of users \emph{relative} to others, we will
rescale the benefits from the reputation by the largest value $X_{\mathrm{max}}(t)$, which makes different networks more
comparable. Hence, the benefits and costs are specified in this model as follows:
\begin{equation}
  \label{eq:benefit}
  b_{i}(t)=\dfrac{X_{i}(t)}{X_{\mathrm{max}}(t)} \in (0,1) \;; \quad
  c_{i}(t)=\tau \in [0,1)
\end{equation}
Note that, in contrast to Eq. \eqref{eq:benefit}, it is common to express the relative reputation in terms of the
sum of the individual reputations, $\sum X_{i}(t)$:
\begin{equation}
  \label{eq:benefit-common}
  x_{i}(t) = \dfrac{X_{i}(t)}{\sum_{j}X_{j}(t)}
\end{equation}

In the context of an OSN,
however, this is unrealistic, as it requires that either the total
reputation or everyone's reputation is public knowledge. For this reason,
we posit that users compare their reputation
to the most reputable individual, who is often visible in rankings of user reputation. 
We note that
despite the conceptual difference, a solution to Eq. \eqref{eq:benefit} can
be mapped directly to a corresponding solution to
Eq. \ref{eq:benefit-common}, by normalising $x_{i}$ with respect to
$x_{\text{max}}$ as follows:
\begin{equation}
  \label{eq:dummy}
  \dfrac{x_{i}(t)}{x_{\text{max}}(t)} = \dfrac{X_{i}(t) /
    \sum_{j}X_{j}(t)}{X_{\text{max}}(t) / \sum_{j}X_{j}(t)} =
  \dfrac{X_{i}(t)}{X_{\text{max}}(t)} = b_{i}(t)
\end{equation}

In Appendix A, we further show that an equilibrium solution to Eq. \eqref{eq:dyn} is
also an equilibrium for $b_{i}(t)$ and $x_{i}(t)$ up to a scaling
factor. This means that the entry/exit dynamics introduced in Section \ref{sec:network} is invariant to the particular way in which users evaluate their relative reputation.

According to Eq. (\ref{eq:stay}), users leave the OSN at time $t$ if their \emph{relative reputation} is lower than a fixed
threshold. Their links $a_{ij}(t)$ are then set to zero, which according to Eq. (\ref{eq:dyn}) reduces the reputation of other
users $j$ at the next time step. This can lead to cascades of users leaving the OSN at consecutive time steps.  
The aim of our paper is to understand how a decrease of users' motivation to stay because of an increase in their fixed costs $\tau$ will impact the OSN.
Therefore, as a next step, in Sect. \ref{sec:equilibrium} we first investigate how the reputation
depends on the social network, before turning to the entry/exit dynamics in Sect. \ref{sec:network}.

\subsection{Quasistationary Equilibrium}
\label{sec:equilibrium}

Let us first discuss the reputation dynamics of users for a \emph{fixed}
social network. Expressing the dynamics of $b_{i}(t)$ from
Eq. \eqref{eq:benefit} yields (see Appendix A):
\begin{equation}
\label{eq:rel-reputation}
\dfrac{db_{i}}{dt} = \sum_{j=1}^{N}a_{ij}b_{j}(t) - b_{i}(t)\sum_{j=1}^{N}a_{zj}b_{j}(t)
\end{equation}
where $z$ is the index of the individual with highest absolute reputation
$X_{\text{max}}$. The first term describes the reputation boost that
individual $i$ obtains from all her followers. The second term is a
scaling factor and represents the reputation decay with strength equal to the total boost in reputation
that user $z$ receives.

\paragraph{Eigenvalues and eigenvectors}

\begin{figure}[htbp]
  \begin{minipage}{0.4\textwidth}
  \centering 
  $\mathbf{A}$ = $\begin{pmatrix}
0 & 1 & 1 & 0 & 0 \\ 
1 & 0 & 0 & 0 & 0 \\ 
0 & 1 & 0 & 0 & 0 \\
0 & 1 & 0 & 0 & 0 \\
0 & 0 & 0 & 1 & 0
\end{pmatrix}$
\end{minipage} 
    \begin{minipage}{0.5\textwidth}
\includegraphics[width=0.99\textwidth]{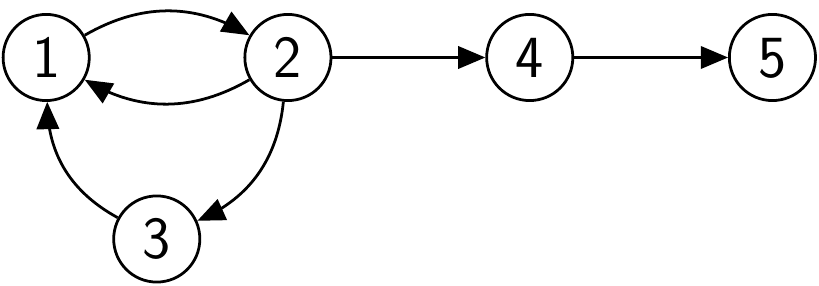}
\end{minipage} 
  \caption{Sketch of an OSN (right) that displays a core-periphery structure (see text) and the corresponding adjacency matrix $\mathbf{A}$ (left)}
  \label{fig:lambda5}
\end{figure}

The set of Eqs. (\ref{eq:rel-reputation}) forms a
linear dynamical system of coupled first-order differential equations
with the initial conditions described by the vector
$\mathbf{b}(0)=\{b_{1}(0),b_{2}(0),...,b_{N}(0)\}$. The 
(constant in time) $a_{ij}$ forms the elements of an adjacency matrix
$\mathbf{A}$ of size $N\times N$ (see Figure \ref{fig:lambda5}). 

At equilibrium we require $\dot{b_{i}}=0$, hence
\begin{equation}
\sum_{j=1}^{N}a_{ij}b_{j}(t) = b_{i}(t)\sum_{j=1}^{N}a_{zj}b_{j}(t)
\end{equation}
In matrix form, this equation becomes
\begin{equation}
\label{eq:rel-eigen}
\mathbf{A} \mathbf{b}(t) = \mathbf{b}(t)\sum_{j=1}^{N}a_{zj}b_{j}(t)
\end{equation}

If $\mathbf{X^{\lambda}}$ is an eigenvector of $\mathbf{A}$ with a
corresponding eigenvalue $\lambda$, then rescaling $\mathbf{X^{\lambda}}$ will
also produce an eigenvector, i.e. $\mathbf{b}^{\lambda} = \mathbf{X^{\lambda}}
/ X_{\text{max}}^{\lambda}$ is a solution to Eq. \eqref{eq:rel-eigen}. In
this case the scaling factor $\sum_{j}a_{zj}b_{j}(t)$ gives the
$m^{\text{th}}$ component, $(\lambda b^{\lambda})_{m}$, of the vector $\lambda
\mathbf{b}^{\lambda}$. Since
$(\lambda b^{\lambda})_{m}=X_{m}^{\lambda} / X_{\text{max}}^{\lambda} =
1$, it follows that $\sum_{j}a_{zj}b_{j}(t) = \lambda$.

Since $\mathbf{A}$ is a real nonnegative matrix, the
\textit{Perron-Frobenius Theorem} tells us that the 
largest eigenvalue of $\mathbf{A}$, denoted by $\lambda_1$, is real and
positive \citep{Berman1979, bapat1997nonnegative}. 
Furthermore, the corresponding Perron-Frobenius eigenvector
$\mathbf{b}^{\lambda_{1}}=\{b_{1}^{\lambda_{1}},b_{2}^{\lambda_{1}},...,b_{N}^{\lambda_{1}}\}$,
is the only eigenvector with 
purely positive entries,
and
defines the reputation values of each user in equilibrium.

Let us illustrate this by the didactical example of a rather small
network shown in Figure \ref{fig:lambda5}.  

The characteristic polynomial determining the
eigenvalues of the corresponding adjacency matrix $\mathbf{A}$ given in Figure \ref{fig:lambda5} is
\begin{equation}
-\lambda^{5} + \lambda^{3} + \lambda^{2}  = 0
\end{equation}
and the largest eigenvalue is $\lambda_1$$=1.32$. The corresponding
eigenvector gives us the absolute reputation:
$\mathbf{X}^{\lambda_{1}} = \{2.32, 1.75, 1.32, 1.32, 1\}$. Rescaling this
eigenvector by $1/X^{\lambda_{1}}_{\text{max}}$ gives the relative reputation in equilibrium:
$\mathbf{b}^{\lambda_1}=\{1, 0.75, 0.57, 0.57, 0.43\}$. 

\paragraph{Core-periphery structure}

{
To facilitate the interpretation of the relative reputation values in Figure \ref{fig:lambda5}, let us take a look at
the corresponding network structure. This toy network already
shows a topological feature typical for many social networks, known as a \emph{core-periphery}
structure \citep{Holme2005,Capocci2006,Hojman2008,Vitali2011,Cattani2008,Kumar2010,Corten2012}. Following
\citet{Borgatti2000,Everett2008}, the core is defined as a cohesive
subgroup (e.g. a clique, $n$-clique, $n$-club or $n$-clan, $k$-plex) and the
periphery is everything else. 

Accordingly, and accounting for the
directionality in reputation-based OSNs, we define the \emph{core}, $Q$, as the
\emph{largest strongly-connected component} (SCC) in the OSN. Each node in the SCC is
reachable from all other nodes in the SCC. The \emph{periphery} consists of all
nodes that do not belong to the core. 
It is known that the Perron-Frobenius eigenvalue $\lambda_{1}$ of
a directed OSN  is equal to the Perron-Frobenius eigenvalue of
its largest SCC and can be used to characterise the structure of the core
$Q$ (see Appendix B). 
In Figure~\ref{fig:lambda5}, the
core is the SCC formed by users \emph{1}, \emph{2}, \emph{3}, who
mutually 
boost each others' reputations. The periphery consists of users \emph{4} and \emph{5}
who only benefit from being connected to the core. The periphery usually
contains simple \emph{chains} of users (\emph{4} $\to$ \emph{5}) which emanate from the
core, in this case from user \emph{2}. The core itself does not contain simple
chains, but \emph{cycles}, i.e. \emph{closed} directed chains that
involve 
$n\geq 2$ users \citep{Jain2002}. In the example, we observe two cycles,
\emph{1} $\to$ \emph{2} $\to$ \emph{1} and \emph{1} $\to$ \emph{2} $\to$
\emph{3} $\to$ \emph{1}, i.e. users can be part of more than one
cycle. Moreover, all users have one follower each, except for user
\emph{1} who has two 
followers, thus her reputation can be expected to be higher than that of
the others. Note that even though users \emph{4} and \emph{5} have the same number
of followers, their reputation is different, as user \emph{4} has a more
reputable follower belonging to the core. 

More formally, a \textit{cycle} is defined as a subgraph in which there is a closed path from every node of the
subgraph back to itself. Cycles and structures of interlocking cycles
represent \textit{irreducible subgraphs}. The core $Q$ must always contain at
least one cycle to qualify as a strongly connected component. In Appendix B we show the dependency between the Perron-Frobenius eigenvalue and the
length and number of the cycles in the core.
}

\paragraph{Direct and indirect reciprocity}
The shortest
cycle, $n=2$, expresses \emph{direct reciprocity}, i.e. in Figure \ref{fig:lambda5} users \emph{1} and \emph{2} mutually follow each other and boost each
other's reputations. For $n\geq 3$, we have \emph{indirect reciprocity}, i.e. user \emph{2} follows \emph{3}, however, \emph{3} does not follow \emph{2},
but follows \emph{1} instead, and only \emph{1} may follow \emph{2}, thus closing the loop.

Direct reciprocity is very common in OSN, e.g. in \texttt{twitter} or \texttt{google+} it is seen as good practice to link back
to someone that has choosen to follow you or to have you as his/her
friend. Likewise, \texttt{likes}, \texttt{+1}, or shared
posts often take reciprocity into account. Compared to this, indirect
reciprocity is more difficult to detect.To boost
interaction along a chain of followers, and to hopefully close the loop,
some OSN, e.g. \texttt{google+} or \texttt{researchgate}, indicate for
each follower the number of additional users that the user and the follower both have in common. This may increase the likelihood of creating
shortcuts and also shorter cycles.

\paragraph{Length of simple chains}

At equilibrium, $db_{i}/dt=0$, we can insert the eigenvector $\mathbf{b}^{\lambda_1}$
corresponding to $\lambda_1$ into Eq. \eqref{eq:rel-reputation} to
get
\begin{equation}
  \sum_{j} a_{ij} b_j^{\lambda_{1}}=\lambda_1 b_i^{\lambda_{1}}\;; \quad 
  b_i^{\lambda_{1}}=\dfrac{1}{\lambda_1}\sum_{j} a_{ij} b_j^{\lambda_{1}}
  \label{attenuation2}
\end{equation} 
This means that, in the long run, the reputation $b_i$ of user $i$ is
equal to the sum of the reputations of all users $j$ that follow $i$, \textit{attenuated} by a factor ${1}/{\lambda_1}$ \citep{Jain1998, Jain2002}.

Eq. \eqref{attenuation2} allows us to draw some conclusions about the
maximum length of simple chains involving peripheral users. In the
example discussed, we  note that the 
reputations of users \emph{4} and \emph{5} are related by the attenuation factor in
Eq. \eqref{attenuation2} such that $b_{4} = b_{2} / \lambda_{1}$, $b_{5} =
b_{2} / \lambda_{1}^{2}$, and in general $b_{n} = b_{2} /
\lambda_{1}^{n-1}$ for a chain of length $n$. 
If we require a simple
chain to be exactly of length $n$, $b_{n} > \tau$ and $b_{n+1} \le \tau$ must hold. In other
words, the $n^{th}$ peripheral user finds it beneficial to stay
while the $(n+1)^{th}$ leaves. Hence, we obtain for $n$:
\begin{equation}
  \label{eq:maxlength-not-simplified}
  n \in \left[\dfrac{\ln (b_{2}/\tau)}{\ln \lambda_{1}}, \dfrac{\ln (b_{2}/\tau)}{\ln
      \lambda_{1}} + 1\right) \;;\quad
  n = \ceil*{\dfrac{\ln (b_{2}/\tau)}{\ln \lambda_{1}}}
\end{equation}
since we require $n$ to be an integer value.
The maximum length of a simple chain in a core-periphery network thus
depends on the cost level $\tau$, the connectedness within the core $Q$
expressed by the largest eigenvalue $\lambda_{1}$, and the relative
reputation of the core user who connects the core to the chain (in our example user \emph{2}).

\paragraph{Unstable cores}

Imposing the condition $n>0$ in Eq. \eqref{eq:maxlength-not-simplified} requires that
$\lambda_{1} > 1$ and $b_{2} > \tau$ which  holds only if the OSN contains 
\emph{cycles}. Without these cycles, $\lambda_{1}=0$, i.e. the core-periphery structure breaks down. The condition, $b_{2} > \tau$, requires that the core user \emph{2}, who connects to the simple chain, needs to obtain a net gain from participating in
the OSN. Otherwise, due to the attenuation factor in
Eq. \eqref{attenuation2}, none of the users in the chain would have
an incentive to stay in the network.

The special case of $\lambda_{1}=1$ represents an
important exception. In this case, and provided that $b_{2}>\tau$,  the
length of the simple chain is no longer bounded, since all users in the
chain would have the same relative reputation. More importantly, however,
network structures like these are very unstable, because core users have
a reputation comparable to the periphery. In fact, peripheral users can
often obtain a higher reputation from multiple connections to the
core, which reduces the benefit of the core users and increases their likelihood of leaving. 
Different from peripheral users, the leave of core users considerably affects other core users that are part of the same cycle. This starts cascades of users  leaving and thus destroys the
core. Without the core, the periphery would not be able to sustain its
reputation and would break down as well.

\paragraph{Number and length of cycles}

Unstable core-periphery structures can be avoided as long as \emph{interlocking cycles} appear in the core. These contain users  involved in multiple cycles which in turn receive a much higher reputation and increase the benefit, i.e. the relative reputation, also for others. 
Both the number of cycles in a network and their length have an impact on
the largest eigenvalue $\lambda_{1}$ as illustrated in Appendix B. In
general, we can conclude that $\lambda_{1}$ \emph{increases} with the
\emph{number of cycles}, but 
\emph{decreases} with the \emph{length} of the cycle (keeping all other variables constant). 

The number of cycles in the network
further depends on the average density $m$ (average number of links per user), a parameter discussed in the next section 
when we introduce the dynamics for the network.

\subsection{Network dynamics}
\label{sec:network}

In the previous section, we have explained that the reputation dynamics of Eq. (\ref{eq:dyn}), for a \emph{fixed network}, converges to an equilibrium state in which the relative reputations of users are fixed. This convergence time defines the time scale for the reputation dynamics. Dependent on their stationary reputation value that defines their benefit in relation to their costs $\tau$, Eq. (\ref{eq:benefit}), users can decide to leave the OSN, Eq. (\ref{eq:stay}), and will be replaced by new users joining the OSN. As described above, we assume that  $N_{\mathrm{exit}}(T)=N_{\mathrm{entry}}(T)$, to keep $N$ constant. $T$ is the time at which entry and exit happen. We assume that the time scale for entry and exit, i.e. for changing the network structure, can be separated from the time scale of the reputation dynamics which is much shorter. I.e. users make their decision based on the quasi-stationary benefit, which can only change after the network has changed. 

If a user leaves the OSN, all her (incoming and outgoing) links are set
to zero. The newcomer will establish the same number of incoming and
outgoing links \emph{on average} but, assuming that she does not know all other users, these links will be distributed at random. We assume that each user follows on average $m$ other users. Precisely, there is a constant probability $p$ that a new user links to the $(N-1)$ other users, and $m=p(N-1)$ is a constant related to  the average density of a random OSN. 

Our major interest in this paper is in the role of the cost $\tau$ that, according to Eqs. (\ref{eq:stay}), (\ref{eq:benefit}) defines the level at which users will leave the OSN, measured in terms of  the relative reputation. The latter is between zero and one, so $\tau$ gives the fractional benefit that has to be reached to stay in the OSN. In our computer simulations, we will vary this level from zero to 0.5 to study the impact of increasing costs. 

$\tau$=0 would imply no costs. To still allow for a network dynamics in this case, we apply the so-called extremal dynamics. I.e. we choose the user with the lowest relative reputation, and force her to leave the OSN, to allow a new user to enter. In case of several users with the same low $b$ value, we choose one of them randomly. The other limiting case $\tau=1$ would imply that all $N$ users will leave and be replaced by a completely new cohort. Then, the network at every time $T$ starts as a new random network, which has no chance to evolve. Hence, small or intermediate values of $\tau$ would be most appropriate.

\section{Results of computer simulations}
\label{sec:simulat}

\subsection{Performance}
\label{sec:perform}

How should one measure the ``performance'' of a OSN? 
Users join the OSN for a purpose and, as we have explained in Sect. \ref{sec:costs}, here we assume that the benefits of users can be measured in terms of their \emph{relative reputation}, which should be possibly increased. 
This implies that not only the number of followers is taken into account, but also their ``value'' in terms of their own reputation. 
Hence, it would be obvious to use the \emph{long-term average} over all users' relative reputation as a systemic measure whether or not the OSN fulfills the expectation of its users. 
This will serve as a proxy for the performance of the OSN.

Specifically, we build on the relative reputation, $b_{i}(T)$, Eq. (\ref{eq:benefit}) of a user $i$, obtained in the
quasi-stationary limit at time $T$. Then, both $X_{i}(t)$ and
$X_{\mathrm{max}}(t)$ can be expressed by the corresponding values
$X_{i}^{\lambda_{1}}(T)$ and $X_{\mathrm{max}}^{\lambda_{1}}(T)$ from the
eigenvector $\mathbf{X}^{\lambda_{1}}$ and we find for the average
reputation of all $N$ users 
\begin{equation}
  \label{eq:satisf}
  \bar{b}(T)=\frac{1}{N}\sum_{i=1}^{N}b_{i}(T)= \frac{1}{N}\sum_{i=1}^{N}\frac{X_{i}^{\lambda_{1}}(T)}{X_{\mathrm{max}}^{\lambda_{1}}(T)} \;;\quad
\mean{b}=\frac{1}{R}\sum_{r=1}^{R}\bar{b}_{r}(T_{\mathrm{max}})
\end{equation}
$\bar{b}(T)$ refers to the population average at a given time $T$ which can considerably fluctuate because of stochastic influences when changing the network structure at every time step $T$. 
Therefore, we define the \emph{long-term average benefit} $\mean{b}$ which is a system average taken over a large number of independent runs $R$, where the average satisfaction was measured after considerable long time $T_{\mathrm{max}}$. 

\begin{figure}[htbp]
  \centering
  \includegraphics[width=0.45\textwidth]{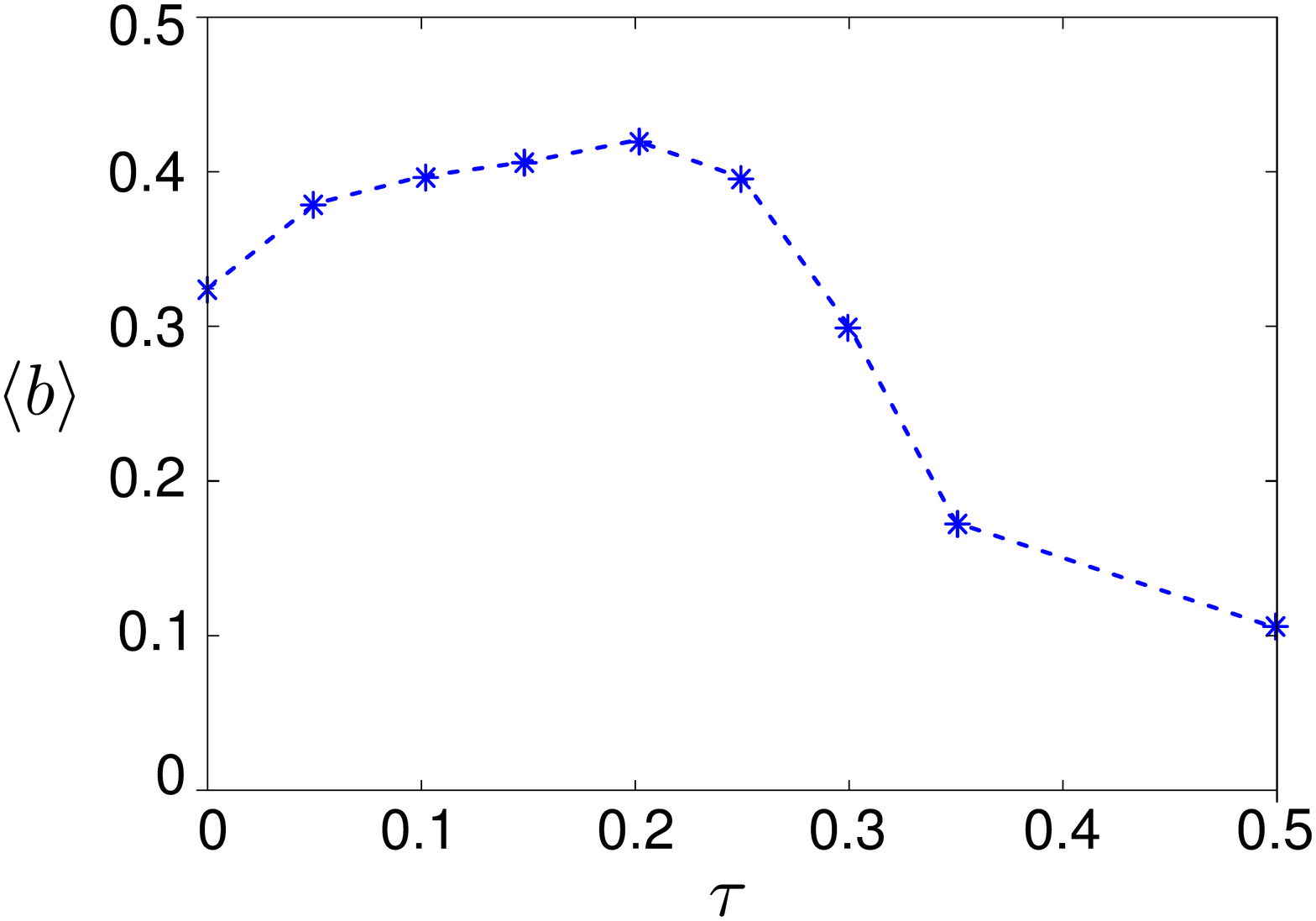}
\hfill
  \includegraphics[width=0.45\textwidth]{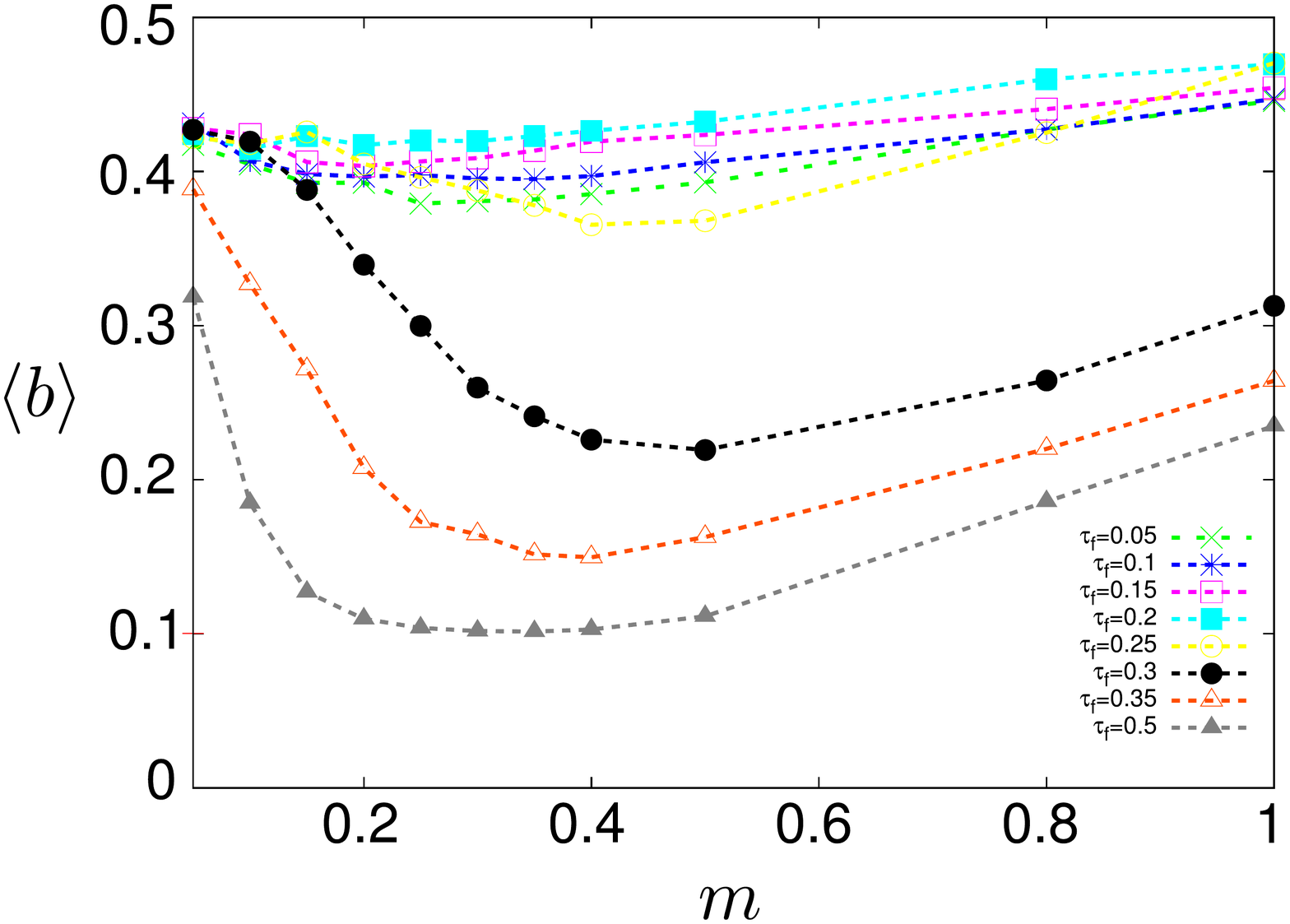}
  \caption{Long-term average benefit $\mean{b}$, Eqn. (\ref{eq:satisf}) for varying costs $\tau$ (left) and average number of links $m$ (right).}
  \label{fig:ff}
\end{figure}

The results of our simulations are shown in Fig. \ref{fig:ff} for various values of the costs $\tau$ and the average number of links $m$. 
The most remarkable observation, prominently shown on the left side of
Figure \ref{fig:ff}, is the maximum average benefit
$\mean{b}_{\mathrm{max}}$ for a non-zero cost at
$\tau_{\mathrm{max}}$=0.2. Counterintuitively, this result implies that adding a cost for participating in an OSN maximises the
average reputation in the system. 

The right part of Figure \ref{fig:ff} demonstrates how this effect depends on the other important model parameter, the average number $m$ of users, a new user tends to follow. Here, we clearly observe that for costs below $\tau_{\mathrm{max}}$, $m$ has almost no influence on the performance, which is interesting enough because one would assume that a larger number of potential followers would always improve the situation. Note, however, that the performance relates to the \emph{relative} reputation, i.e. $X_{i}$ may increase with $m$ but so does $X_{\mathrm{max}}$. For costs above $\tau_{\mathrm{max}}$, we see a drastic decrease in the performance which depends on $m$ in a much more pronounced way. In fact, there is a non-monotonous dependence, i.e. \emph{increasing} $m$ in the range of small values will further \emph{decrease} the performance. 

To better understand these interesting results about the improvement of performance with increasing costs, we have to refer to the robustness of the OSN, which is inherently related to the stability of the core-periphery structure already discussed in Sect. \ref{sec:equilibrium}.

\subsection{Robustness}
\label{sec:robustness}

A OSN is said to be \emph{robust} if the decision of users to leave the OSN will not trigger  large cascades of further users leaving. 
Hence, we could use the sheer fraction of users remaining in the OSN at
time $T$, $Y(T)=1-N_{\mathrm{exit}}(T)/N$, as a measure of its robustness. However, because this dropout is always compensated by a
number of new users entering the OSN, we will need a different robustness measure.

We recall that the robustness of the OSN depends  on the existence of a distinct \emph{core-periphery} structure for which the stability conditions are explained in Sect. \ref{sec:equilibrium} and Appendix C. 
This core-periphery structure is challenged at every time $T$ because of the entry and exit of users and the corresponding formation and deletion of links. 
Therefore, cascades of users leaving can generally not be avoided, as  exemplified in Appendix C. 
However, such cascades should not destroy the whole OSN, in particular not the \emph{core} of the network that contributes the most to the average benefit of its users and prevents the periphery from leaving the OSN. 

Consequently, we will use the \emph{life time} $\Omega_{Q}$ of the core $Q$ (measured in network time $T$) as a proxy for the robustness of the OSN. 
Because of considerable fluctuations, similar to performance, we will use the \emph{average} $\mean{\Omega_{Q}}$ taken over a considerable long time $T_{\mathrm{max}}$ (and possibly also averaged over a large number $R$ of independent runs). 

Figure \ref{fig:mean-core-recovery} in Appendix D illustrates, for a sample network, that the robustness measure $\mean{\Omega_{Q}}$ is maximized for the optimal cost level $\tau_{\mathrm{max}}$=0.2. 
Specifically, we notice a non-monotonous behavior. 
The lifetime of the core-periphery structure is mostly threaten by peripheral users attracting followers, without reciprocally contributing to the benefit of others (see Appendix C). A non-zero cost, $0<\tau<\tau_{\mathrm{max}}$, prevents this ``behavior'' to some extent, but a cost too high, $\tau_{\mathrm{max}}<\tau<1$, rather destroys the core and, hence, the depending periphery. 

Because of the fact that both performance and robustness are maximized for the optimal cost level $\tau_{\mathrm{max}}$=0.2, 
one could argue that performance is simply a substitute of robustness. 
This simplified explanation, however, does not hold. 
As Appendix C shows in detail, the core can be destroyed by cascades that involve core users. 
But the replacement of users leaving by new users at each time step $T$ also bears the chance that random rewiring leads to the \emph{recovery} of the core, thanks to new users linked to the core.

That means, to maximize performance not only the \emph{life time} of the core matters, but also the \emph{recovery time} $\Pi_{Q}$ it takes to rebuild the core. 
The OSN should recover quickly, to not affect the long-term benefit of its users.
As Figure \ref{fig:mean-core-recovery} in Appendix D shows for a sample network, 
the averaged recovery time $\mean{\Pi_{Q}}$ of the core-periphery structure becomes shorter if $\tau$ increases, i.e. if more new users enter the OSN, and new links increase the chance of establishing reciprocal relations. But, again, if the randomness associated with this process becomes too high, favorable structures may get destroyed. A small, but considerable cost $\tau_{\mathrm{max}}$=0.2 is able to balance these counteracting processes. 

In conclusion, the cost value $\tau_{\mathrm{max}}$=0.2 optimizes the \emph{ratio} between these two time spans, the average life time of the core, $\mean{\Omega_{Q}}$, and the average recovery time of the core, $\mean{\Pi_{Q}}$. 
Their combined impact, i.e. maximizing $\mean{\Omega_{Q}}$ while minimizing $\mean{\Pi_{Q}}$ only explains the maximum of performance $\mean{b}$, as discussed in detail in Appendix D.

\subsection{Core size and largest eigenvalue}
\label{sec:core}

\begin{figure}[htbp]
  \centering
  \includegraphics[width=0.45\textwidth]{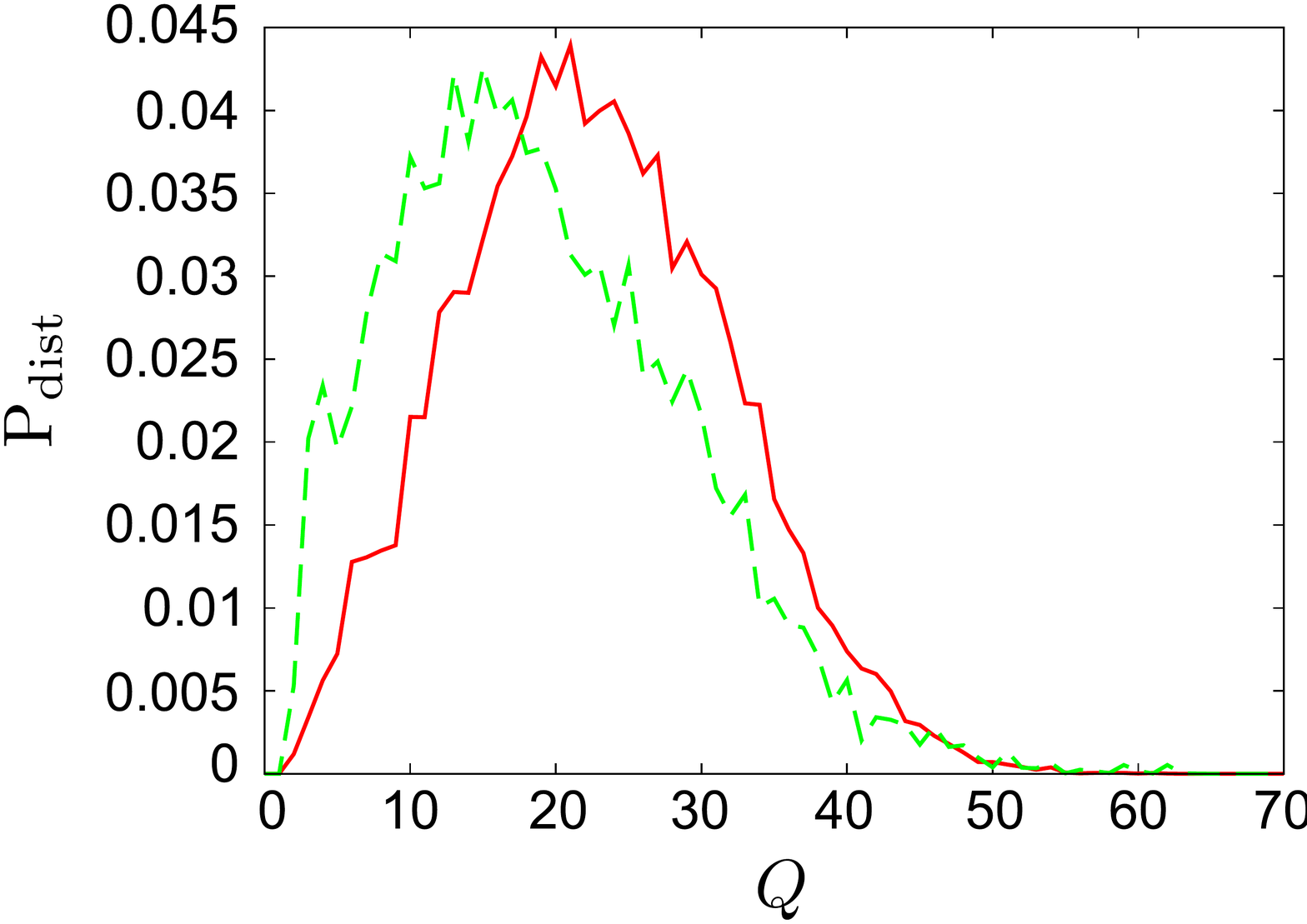}\hfill
  \includegraphics[width=0.45\textwidth]{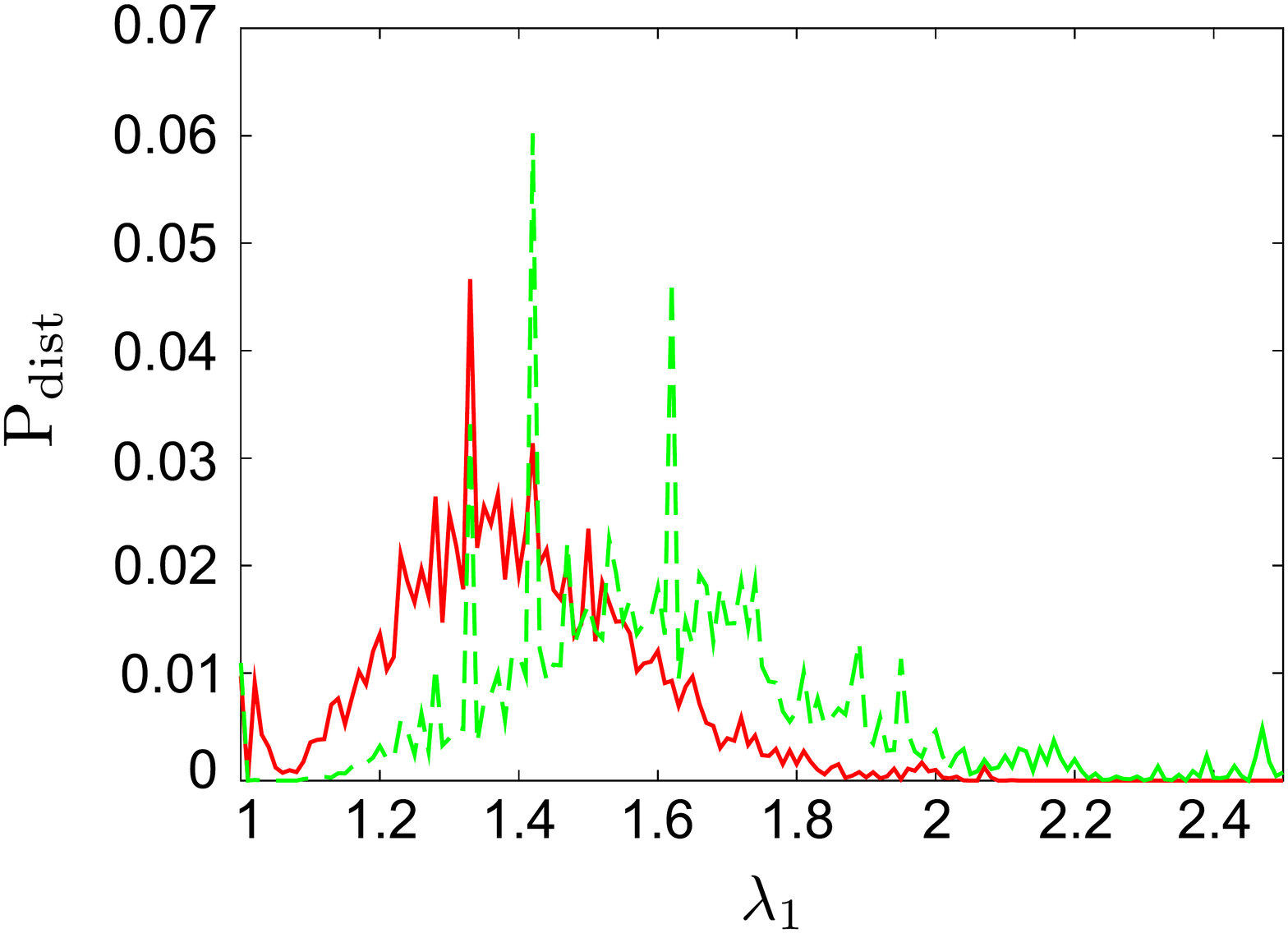}
\hfill
  \caption{(left) Distribution $P(Q)$ of core sizes $Q$, (right)
    Distribution $P(\lambda_{1})$ of the largest eigenvalue
    $\lambda_{1}$. (red) $\tau=0$, (green $\tau=0.25$). Other parameters
    $N=100$, $m=0.25$, $T_{\mathrm{max}}=10^{6}$}
  \label{fig:core}
\end{figure}

In Sect. \ref{sec:equilibrium} we already shortly discussed how an
increase in cost $\tau$ affects the structure of the OSN. In particular,
the length of both simple chains and of cycles of followers will be
reduced (Eq. \eqref{eq:maxlength-not-simplified}). This results in a decreasing size $Q$ of the core built by users that belong to one or more cycles. Figure \ref{fig:core} shows the histogram of the core sizes, $P(Q)$ for two different costs, $\tau$=0 and $\tau$=0.25, where the first one is only used as a reference case. One recognizes that, with increasing cost, the distribution gets more skewed, with its maximum shifted to smaller values of $Q$. To allow for a real comparison of the different simulations, we have taken into account only those realizations where the core-periphery structure encompasses the whole network (i.e. one connected component, and no isolated users, or groups of users). Figure \ref{fig:networks} shown in Appendix E also illustrates the structure of the resulting OSN. One clearly observes that with increasing costs there is a tendency of users in the core to follow more other users.

The second structural insight comes from the histogram of the largest
eigenvalue, $P(\lambda_{1})$, shown in the right part of
Figure \ref{fig:core}. Compared to the case with no costs, we observe a
considerably broader distribution, with the maximum shifted to a larger
value of $\lambda_{1}$. I.e. with increasing costs, there is a much
larger likelihood to find larger values of $\lambda_{1}$, which implies
shorter simple chains and shorter cycles. In line with the argumentation
in Appendix B, we can also confirm that larger $\lambda_{1}$ correspond
to more cycles inside the core, and thus larger average reputation, which can be verified by comparing the snapshots of Figure \ref{fig:networks}. 

We remind that the condition (\ref{eq:maxlength-not-simplified}) only refers to simple cycles and \textit{simple} chains, that is, chains in
which each user has exactly one follower. If users have several followers, their benefit-to-cost difference can overcome the
condition (\ref{eq:maxlength-not-simplified}) even if their position in the chain or in the cycle would not allow for this. And such users
automatically boost the reputation of all other users downstream. As the cost $\tau$ increases, having more than one follower
becomes crucial in particular for peripheric users to stay.  Having more than one follower, on the other hand, also increases the
chance of creating new cycles, which in turn increases $\lambda_{1}$. This again feeds back both on the core and the periphery of
the OSN, increasing the pressure towards a more compact core and shorter chains. Hence, it is in fact the relation with $\lambda_{1}$ that facilitates the role of  the cost $\tau$ in shaping the network.

\section{Discussion}
\label{sec:disc}

In this paper, we have proposed an agent-based model of online social networks (OSN) which captures several generic features of such networks:

\paragraph{Core-periphery structure}
Most OSNs are characterised by a core
of active users that are closely linked to each other, and a periphery of
less active users that are loosely connected to the core. 
Empirical studies have shown \citep{Garcia2013b,Mavrodiev2016} that the core of OSN is not only quite large, compared to the size of the OSN, but also ``deep'', i.e. most users are well integrated in the OSN. 
Our model is able to reproduce such a structure, but also to explain its origin, as a combination
of direct and indirect reciprocity. 
The former implies mutual directed links between two users, whereas the latter occurs if users are part of closed cycles and thus benefit from others through the cycle.

\paragraph{Reputation dynamics} 

Many models of OSN take the number of ``friends'', i.e. direct links between users into accout and take  
degree and traditional centrality metrics as proxies of a user's reputation \citep{Kwak2010,Magno2014,Niven2015}. 
Our model, in contrast, not only considers that such social relations are \emph{directional}, i.e. a link from user $j$ to $i$ does not necessarily implies a link from $i$ to $j$. 
It also explicitely takes the different impact of users into account, weighting it by means of a (generalized) \emph{reputation}.
Noteworthy, reputation is not determined by the user herself, instead it results from the interaction with others. 
A user is said to have a high reputation if it is followed by many other users of high reputation. 
I.e. in addition to the \emph{number} of followers, also their \emph{value} is taken into account. 
Reputation that is not actively maintained will fade out in the course of time. 
I.e. users that are not able to attract followers automatically loose in impact.

\paragraph{Cost-benefit analysis }

In our model, users join the OSN for a purpose. That means they bear costs, to obtain some benefit. 
If their costs are higher than their benefits, users \emph{leave} the OSN. 
Hence, our model includes an \emph{entry-exit dynamics} which is motivated by rational arguments. 
We have assumed that a user's benefit increases with the number of followers it attracted, which in turn boost its reputation. 
Hence, benefits are measured by the relative reputation of a user, scaled by the maximum reputation in the OSN. 
The latter allows us to better compare OSNs of different sizes and link densities. 
Costs, on the other hand, are assumed to be fixed, they include for example the effort of using the website or to maintain a profile.

Our main focus in this paper was on the role of the \emph{cost} $\tau$ incurred for every active users. 
Increasing $\tau$ would imply a pressure on users to leave because it worsens their cost-benefit relation. 
This short-term negative effect, however, has a positive influence on the core-periphery structure of the OSN.
As we demonstrate in this paper, an optimal cost level is able to increase the \emph{long-term benefit} of users in terms of a higher relative reputation.

\paragraph{Performance}

In order to extimate the impact of increasing costs, we have defined the long-term benefit averaged over the whole OSN, $\mean{b}$, as a performance measure. Intuition would  suggest that $\mean{b}$ monotonously decreases with increasing costs up to a point where the whole OSN collapses. Interestingly, this picture does not hold for comparably small cost values. On the contrary, a small cost up to $\tau_{\mathrm{max}}$=0.2 improves the situation, i.e. the long-term benefit \emph{increases} compared to a reference case without any cost (where no user would leave). In other words, a small cost forces those users to leave which were never able to attract any follower. This, in our model, gives way to new users that might be more successful in this respect. 

\paragraph{Robustness}

As a second important insight, we analyze both mathematically and by means of computer simulations  how increasing costs change the structure, and hence the robustness, of the OSN. 
Our model is set up in a way that it allows to express the outcome of the reputation dynamics by $\lambda_{1}$, the largest eigenvalue of the adjacency matrix that describes the interaction between users, i.e. who follows whom. 
We verified that an increasing cost $\tau$ leads to higher values of $\lambda_{1}$, which means a smaller, but more compact core
and, most remarkably, in an increasing likelihood to have more than one follower.  
This, on the other hand, \emph{increases} the direct and indirect \emph{reciprocity} that characterizes the core. 

Robustness implies that cascades of users leaving will not destroy the core, i.e. the life time of the core can be used as a proxy measure of robustness. 
We have shown that this life time is maximized for the optimal cost value $\tau_{\mathrm{max}}$=0.2.

\bigskip

We emphasize that the above conclusions are obtained from a model that, as every model, only captures part of the features of real OSN. 
The value of the model is in what it is able to produce \emph{despite} some of the simplified assumptions. 
Our insights now allow us to specifically search for optimal cost-benefit relations, or to test other assumptions to calculate the benefit.

Why is this issue important? First of all, there \emph{are} costs
involved in being a member of an OSN, even if they are
implicit. These include, among others, the efforts to login and to
retain one's own social network (e.g. by maintaining regular information stream), the effort of learning how to
efficiently use the GUI of the provider, or the effort to adapt
to changes in the GUI. The latter can make it harder to maintain
social contacts, at least temporarily. As \citet{Garcia2013b} argued, it
was ill-timed interface changes that caused the massive dropout of users from
\texttt{Frienster}, which came at a time when new competitors, such as
\texttt{Facebook}, were ramping up in popularity.

There is a second issue involved in this discussion. Even if most users
enjoy participating in an OSN free of charge, companies would like to
know these users would respond, if at some point in time costs
such as monthly membership fees are introduced. Will members remain
loyal, or will this lead to a massive exodus of users, making the OSN
less attractive for investors? Subsequently, if users leave how would
this affect the OSN? What will be the impact of less active users leaving
compared to core users leaving? What incentives should be introduced to keep power users engaged?

With our agent-based modeling framework that consider both entry-exit dynamics and cost-benefit considerations of users, we are able to address such questions. 
It also helps to better understand the relation between the integration of users and the overall performance of the OSN, which allows to formally consider issues such as user satisfaction and service quality.

\bibliographystyle{sg-bibstyle-nourl}
\small \setlength{\bibsep}{1pt}

\normalsize 
\section*{Appendix A: Absolute and relative reputation}
In this Appendix, we show that the equilibrium \textit{absolute} reputations given
by a stationary solution to Eq. \eqref{eq:dyn} are a scaled representation of the
equilibrium \textit{relative} reputations $\mathbf{x}$ and
$\mathbf{b}$. 

Let us rewrite Eq. \eqref{eq:dyn} in matrix form:
\begin{equation}
\label{eq:appendix-dyn}
\dfrac{d\mathbf{X}(t)}{dt} = \mathbf{A}\mathbf{X}(t) - \phi \mathbf{X}(t)
\end{equation}
It is clear that an
equilibrium exists only if $\phi$ is an eigenvalue of
$\mathbf{A}$. In this case, the corresponding eigenvector gives the
equilibrium absolute reputation. Additionally, as $\phi$ represents a
decay rate, and $X_{i}(t) \ge 0$ for all $i$, we require that the matrix $\mathbf{A}$ has at
least one positive eigenvalue, and that its corresponding eigenvector
does not have negative components. As $\mathbf{A}$ is a non-negative real
matrix, we obtain these conditions from the Perron-Frobenius theorem (see
main text). However, we are still bound by a limited set of values for
the decay rate.

To avoid this last restriction, a common transformation is to normalise the
absolute reputation, $X_{i}(t)$, with respect to the total reputation in
the system. By doing so, the free parameter $\phi$ disappears, provided
it is the same for all users. More importantly, using such relative
reputation is more plausible in an OSN where individuals tend to evaluate
their benefit with respect to others, rather than to an arbitrary
absolute scale. 

Let $x_{i}(t) = X_{i}(t) / \sum_{j}X_{j}(t)$. Then we can express the
dynamics of $x_{i}(t)$ as follows:

{\small
  \begin{align}
  \label{eq:appendix-relative-sum}
  \dfrac{dx_{i}(t)}{dt} &=
  \dfrac{\dfrac{dX_{i}(t)}{dt}\left(\sum_{j}X_{j}(t)\right)}{\left(\sum_{j}X_{j}(t)\right)^{2}}\,
  -\,
  \dfrac{X_{i}(t)\dfrac{d\left(\sum_{j}X_{j}(t)\right)}{dt}}{\left(\sum_{j}X_{j}(t)\right)^{2}}
= \dfrac{dX_{i}(t)/dt}{\sum_{j}X_{j}(t)} -
x_{i}(t)\dfrac{d\sum_{j}X_{j}(t)/dt}{\sum_{j}X_{j}(t)} \nonumber  \\[7pt]
&=\dfrac{\sum_{j}
a_{ij} X_j(t) -\phi X_i(t)}{\sum_{j}X_{j}(t)} - x_{i}(t)\dfrac{\sum_{j}\sum_{k}
a_{jk} X_k(t) -\phi X_j(t)}{\sum_{j}X_{j}(t)}  
\nonumber \\[5pt]
&= \sum_{j}a_{ij}x_{j}(t) - \cancel{\phi x_{i}(t)} + \cancel{\phi
  x_{i}(t)} - x_{i}(t)\sum_{j}\sum_{k}a_{jk}x_{k}(t) \nonumber \\ &
                                                                     = \sum_{j}a_{ij}x_{j}(t) - x_{i}(t)\sum_{j}\sum_{k}a_{jk}x_{k}(t)
        \end{align}
        }

        In matrix form Eq. \eqref{eq:appendix-relative-sum} becomes:
\begin{equation}
  \label{eq:appendix-relative-sum-matrix}
  \dfrac{d\mathbf{x}(t)}{dt} = \mathbf{A}\mathbf{x}(t) - \mathbf{x}(t)\sum_{j}\sum_{k}a_{jk}x_{k}(t)
\end{equation}
The equilibrium solutions to Eq. \eqref{eq:appendix-relative-sum-matrix}
are given by the eigenvectors of $\mathbf{A}$. If $\mathbf{X^{\lambda}}$
is one such eigenvector with a corresponding eigenvalue $\lambda$, then the rescaled vector
$\mathbf{x}^{\lambda} = \mathbf{X^{\lambda}} / \sum_{i}X_{i}^{\lambda}$
is also an eigenvector, and the constant,
$\sum_{j}\sum_{k}a_{jk}x_{k}(t)$, equals $\lambda$. 

However, defining relative reputation in this way,
presumes that either everyone's reputation or the total reputation in the
system is common knowledge. This is unlikely to be the case in any
OSN. For this reason, we normalise $X_{i}(t)$ with respect to the individual with
maximum absolute reputation $X_{\text{max}}(t)$.

Let $b_{i}(t) =
X_{i}(t)/X_{\text{max}}(t)$. The dynamics of
$b_{i}(t)$ becomes:

{\small
  \begin{align}
 \label{eq:appendix-max-norm}
\dfrac{db_{i}(t)}{dt} &=
\dfrac{\dfrac{dX_{i}(t)}{dt}X_{\text{max}}(t)}{X_{\text{max}}(t)^{2}} -
\dfrac{X_{i}(t)\dfrac{dX_{\text{max}}(t)}{dt}}{X_{\text{max}}(t)^{2}} =
=\dfrac{dX_{i}(t)}{dt}\dfrac{1}{X_{\text{max}}(t)} -
b_{i}(t)\dfrac{1}{X_{\text{max}}(t)}\dfrac{dX_{\text{max}}(t)}{dt}                           \nonumber
 \\[5pt]
&=\dfrac{\sum_{j}a_{ij}X_{j}(t)}{X_{\text{max}}(t)} -\phi
\dfrac{X_{i}(t)}{X_{\text{max}}(t)} -
b_{i}(t)\dfrac{1}{X_{\text{max}}(t)}\left(\sum_{j}a_{zj}X_{j}(t)- \phi X_{\text{max}}(t)\right) \nonumber \\[5pt]
&=\sum_{j}a_{ij}b_{j}(t) -\cancel{\phi b_{i}(t)}+\cancel{\phi b_{i}(t)} -
b_{i}(t)\sum_{j}a_{zj}b_{j}(t)   =\sum_{j}a_{ij}b_{j}(t) - b_{i}(t)\sum_{j}a_{zj}b_{j}(t)
\end{align}
}

where $z$ is the index of the individual with the highest absolute reputation
 $X_{\text{max}}(t)$. 

In a similar manner, a rescaled eigenvector of $\mathbf{A}$, 
$\mathbf{b}^{\lambda} = \mathbf{X^{\lambda}} / X_{\text{max}}^{\lambda}$,
is an equilibrium solution to Eq. (\ref{eq:appendix-max-norm}). In this
case the constant, $\sum_{j}a_{zj}b_{j}(t)$ again equals $\lambda$.

To summarise, an eigenvector $\mathbf{X^{\lambda}}$ of $\mathbf{A}$ is a
solution to both Eq. \eqref{eq:appendix-relative-sum-matrix} and
Eq. \eqref{eq:appendix-max-norm}. This shows that the two ways of defining
relative information are in fact equivalent, up to a scaling factor.

\section*{Appendix B: Impact of cycles on $\lambda_{1}$}

To further illustrate the impact of the network structure, in particular the number and the length of cycles, on the largest
eigenvalue of the adjacency matrix, we present some didactical examples
in Table \ref{tab:examples}.

{\small
\begin{longtable}[ht]{cc}
    \hline
\textbf{Adjacency matrix} & \textbf{Corresponding network} \\ \hline
  \begin{minipage}{0.01\textwidth}
    {\LARGE [1]}
  \end{minipage}
  \begin{minipage}{0.4\textwidth}
      \centering
\vspace{10pt}
  A = $\begin{pmatrix}0 & 0 & 1 \\ 1 & 0 & 0 \\ 0 & 1 & 0\end{pmatrix}$ 
  \\ \vspace{5pt} $ \lambda_{1}$=$1$ \\ $\mathbf{b}^{\lambda_1}$=$(1,1,1)$      
    \end{minipage} &   
    \begin{minipage}{0.28\textwidth}
      \includegraphics[scale=0.5]{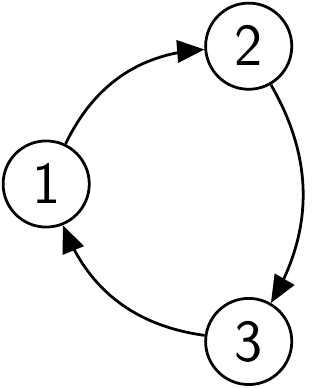} 
\end{minipage} \vspace{5pt} \\ \hline
  \begin{minipage}{0.01\textwidth}
    {\LARGE [2]}
  \end{minipage}
    \begin{minipage}{0.4\textwidth}
      \centering
      \vspace{5pt}
  A = $\begin{pmatrix}0 & 0 & 1 & 0 \\ 1 & 0 & 0 & 0 \\ 0 & 1 & 0 & 1 \\
    0 & 1 & 0 & 0\end{pmatrix}$ 
  \\ \vspace{10pt} $ \lambda_{1}$=$1.22$ \\ $\mathbf{b}^{\lambda_1}$=$(0.819,0.671,1,0.55)$      
    \end{minipage} &   
    \begin{minipage}{0.28\textwidth}
      \includegraphics[scale=0.5]{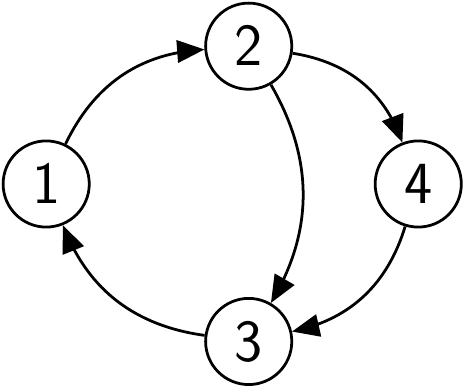} 
\end{minipage}\vspace{5pt} 
  \\ \hline
  \begin{minipage}{0.01\textwidth}
    {\LARGE [3]}
  \end{minipage}
      \begin{minipage}{0.4\textwidth}
      \centering
      \vspace{5pt}
  A = $\begin{pmatrix}0 & 0 & 1 & 0 & 0\\ 1 & 0 & 0 & 0 & 0 \\ 0 & 1 & 0
    & 1 & 0\\
    0 & 1 & 0 & 0 & 1 \\ 0 & 0 & 0 & 1 & 0\end{pmatrix}$ 
  \\ \vspace{5pt} $ \lambda_{1}$=$1.36904$ \\ $\mathbf{b}^{\lambda_1}$=$(0.73,0.533,1,0.835,0.61)$      
    \end{minipage} &   
    \begin{minipage}{0.28\textwidth}
      \includegraphics[scale=0.5]{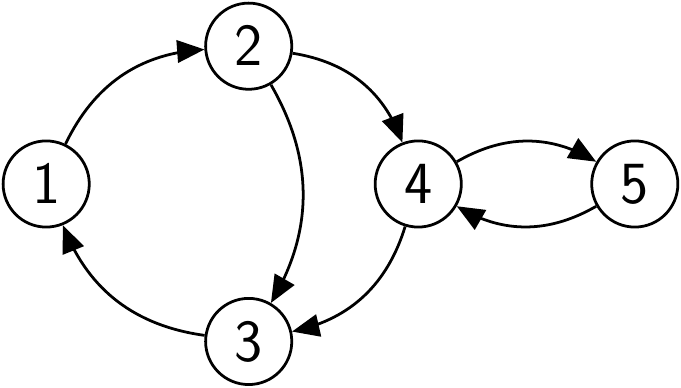} 
\end{minipage} \vspace{5pt} \\ \hline
\begin{minipage}{0.01\textwidth}
    {\LARGE [4]}
  \end{minipage}
      \begin{minipage}{0.4\textwidth}
      \centering
      \vspace{5pt}
  A = $\begin{pmatrix}0 & 0 & 1 & 0 & 0\\ 1 & 0 & 0 & 0 & 0 \\ 0 & 1 & 0
    & 0 & 1\\
    0 & 1 & 0 & 0 & 0 \\ 0 & 0 & 0 & 1 & 0\end{pmatrix}$ 
  \\ \vspace{10pt} $ \lambda_{1}$=$1.194$ \\ $\mathbf{b}^{\lambda_1}$=$(0.838,0.702,1,0.588,0.492)$      
    \end{minipage} &   
    \begin{minipage}{0.28\textwidth}
      \includegraphics[scale=0.5]{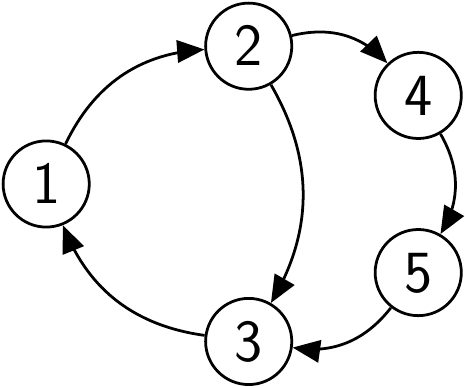} 
\end{minipage} \vspace{5pt} \\
\hline
\begin{minipage}{0.01\textwidth}
    {\LARGE [5]}
  \end{minipage}
      \begin{minipage}{0.4\textwidth}
      \centering
      \vspace{5pt}
  A = $\begin{pmatrix}0 & 0 & 1 & 0 & 0 & 0\\ 1 & 0 & 0 & 0 & 0 & 0\\ 0 & 1 & 0
    & 1 & 0 & 0\\
    0 & 1 & 0 & 0 & 0 & 1 \\ 0 & 0 & 0 & 1 & 0 & 0 \\ 0 &0 & 0 & 0 & 1 & 0\end{pmatrix}$ 
  \\ \vspace{10pt} $ \lambda_{1}$=$1.325$ \\ $\mathbf{b}^{\lambda_1}$=$(0.755,0.57,1,0.755,0.57,0.43)$      
    \end{minipage} &   
    \begin{minipage}{0.28\textwidth}
      \includegraphics[scale=0.5]{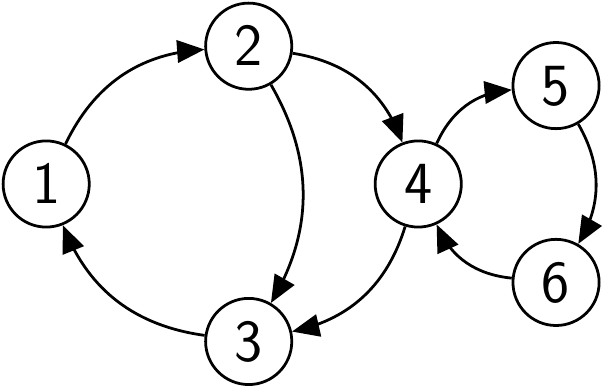} 
\end{minipage}
\vspace{5pt} \\\hline
\caption{Examples of how the network structure affects the adjacency matrix}
  \label{tab:examples}
\end{longtable}
}

The example $[1]$ has the largest eigenvalue of $\lambda_1$$=1$, and
the corresponding eigenvector gives the same  relative reputation for every user. 
A possible next step in the evolution of the core structure is shown in $[2]$
where a new user \emph{4} joins the network. User \emph{2} follows this newcomer, who in turn links itself back to user \emph{3}. The core is now composed of two cycles of length 3 and 4. As a
consequence, $\lambda_1$\ jumps up from $1$ to a new value $1.22$. Hence, the addition of a cycle to the core increases
$\lambda_1$. With this, the reputation of the users in the core is no longer homogeneous. Because of her two followers, user \emph{3}
has the highest relative reputation, wheres user \emph{4}, being the last
in the cycle starting with user \emph{3} has the lowest reputation.

If a new user \emph{5} joins the core structure, this time adding a
2-cycle, as shown in $[3]$, this results in a further increase of $\lambda_1$\ and a
re-shuffling in the reputation vector. User \emph{3} is still dominant, but now user \emph{4} is able to boost her own
reputation thanks to user \emph{5}. This leaves user \emph{2} as the loser in this new configuration.

To illustrate the  the effect of cycle sizes on the core structure,
imagine that user \emph{5} joined the network in a different way. Instead of
creating a 2-cycle, she might have extended an existing one, as shown in $[4]$.
The core now consists of a 3-cycle and a 5-cycle,
everything else being the same (i.e. user \emph{2}  still follows two
other users and user \emph{3} still has  two followers). Yet, this
alternative addition of user \emph{5} has the effect of \textit{reducing}
the value of $\lambda_1$: it is now $\lambda_1$$=1.194$, down from $1.22$ before. User \emph{3} is still dominant, but
now user \emph{5}, being the last in the cycle starting at user \emph{3}, has the lowest population. This core structure would be unfeasible in a
system with a cost level $\tau$$=0.5$, since user \emph{5} would decide to leave.

Let us go back to the sustainable core structure shown in $[3]$. Imagine that
the 2-cycle is expanded by user \emph{6} following user \emph{4} and
being followed by user \emph{5} (shown in $[5]$). Again, the effect of increasing the length of the cycle was to reduce $\lambda_1$.

To summarize these observations we conclude that adding a cycle to an existing core while keeping all else equal increases the
value of $\lambda_1$.  On the other hand, if the number of cycles is kept constant, then \textit{increasing} the
length of a cycle \textit{decreases} the value of $\lambda_1$. 
\section*{Appendix C: Cascades of users leaving the OSN}

Table \ref{tab:core-crash} shows  consecutive steps without cost ($\tau_{1}$=0, left column) and with optimal cost ($\tau_{2}$=0.2, right column), to illustrate how cascades of users leaving the OSN emerge.  
The initial configuration (STEP $T_{0}$) shown in Table \ref{tab:core-crash} has a  core-periphery structure (users \emph{1}, \emph{2}, \emph{3} belong to the core coloured in red, and users \emph{4}, \emph{5} to the periphery).  Users coloured in grey leave the OSN at the given time step. In the initial configuration user \emph{5}  has the
lowest relative reputation of $0.24$  and will leave
the network because, according to Sect. \ref{sec:network}, it is assumed
that even with $b>\tau$ at least one user, i.e. the one with the least
benefit, will leave to ensure the network dynamics for both cost
levels. In the next time step, it will be replaced by another user \emph{5} with
a different connections to the network. Because of the tightly connected
core, most dropout events during subsequent time steps (not shown) will
likely affect only the periphery, replacing users \emph{4} and \emph{5}. However, due
to the random rewiring of $m$ incoming and outgoing links for each new user, configurations like the ones shown at a later STEP $T_{1}$ of Table \ref{tab:core-crash} can appear. In fact, the consecutive steps $T_{1}$-$T_{4}$ are drawn from real simulations of this small OSN (see also Appendix D). 

Interestingly, in STEP $T_{1}$ the peripheral users attract a number of incoming links. This may not happen immediately at their entry, but we assume that, within one time step $T$, after a short relaxation time users become known to at least part of the network and then are able to receive incoming links, as well as they get to know part of the network to establish outgoing links to existing users. 
Because of this link structure, the peripheral users can receive the same relative reputation (printed below each snapshot) as the core users. I.e. all users become susceptible to dropout in the same manner. In the example user \emph{1} (left column) and user \emph{2} (right column) will leave because, among users with the same low reputation, one is chosen at random. Their dropout leads to a smaller core and, consequently, to a lower relative reputation of core users which becomes even lower than the one of peripheral users. This in turn determines further core users to leave as it is shown 
at STEP $T_{2}$. In the right column, user \emph{1}, which was a follower of the previous user \emph{2}, leaves now. In the left column, a new user \emph{1} has entered the OSN, but (because of the random rewiring process) fails to establish links to other users, or to attract other links. Hence, this user is also determined to leave at the current time step. In the consecutive STEP $T_{3}$, the cascade of core users leaving further erodes the structure of the core and hence also affects the peripheral users as shown in STEP $T_{4}$. In conclusion, after the core was affected by the cascades in STEP $T_{1}$, the OSN could no longer remain resilient. 

As the snapshots of Table \ref{tab:core-crash} come from real computer simulations, we can also tell that it took a considerable time before, because of random rewiring, a new core of users emerged. Precisely, the recovery time was  $T_{\mathrm{recovery}}$=492 in the case of $\tau_{1}$=0, but only $T_{\mathrm{recovery}}$=297 in the case of $\tau_{2}$=0.2. As this example indicates, the emergence of cascades of users leaving is largely independent of the precise cost level. However the time for the OSN to recover and the average life time of the core significantly depends on it. This is quantitatively analysed in Appendix D.

{\small
\begin{longtable}[ht]{cc}
\bottomrule[0.1em] 
\multicolumn{2}{c}{
\textbf{STEP} $T_{0}$} 
\\ \toprule[0.05em] \\
\multicolumn{2}{c}{
\begin{minipage}{\textwidth}
  \centering
 \includegraphics[scale=0.5]{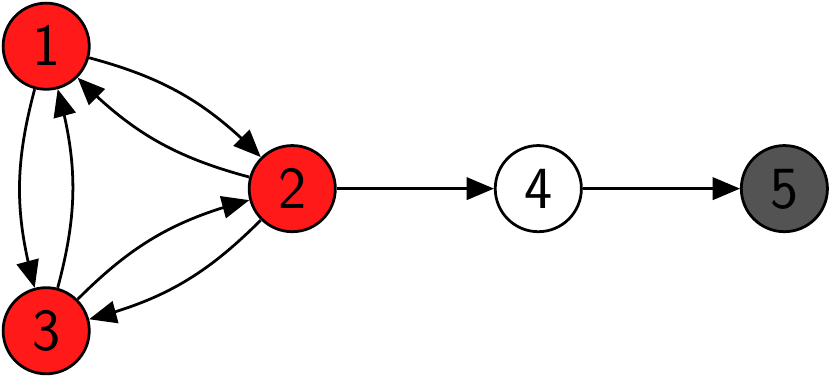}\\ \vspace{5pt}
$\mathbf{b}^{\lambda_1}$=$(1,1,1,0.49,0.24)$\\
 $\langle b \rangle$=$0.74$\vspace{10pt}
\end{minipage}
} \\ 
$\mathbf{\boldsymbol{\tau}_{1}=0}$ & $\mathbf{\boldsymbol{\tau}_{2}=0.2}$ \\ \bottomrule[0.1em]
\multicolumn{2}{c}{\textbf{STEP} $T_{1}$}\\ \toprule[0.05em]
\begin{minipage}{0.5\textwidth}
    \centering
    \includegraphics[scale=0.5]{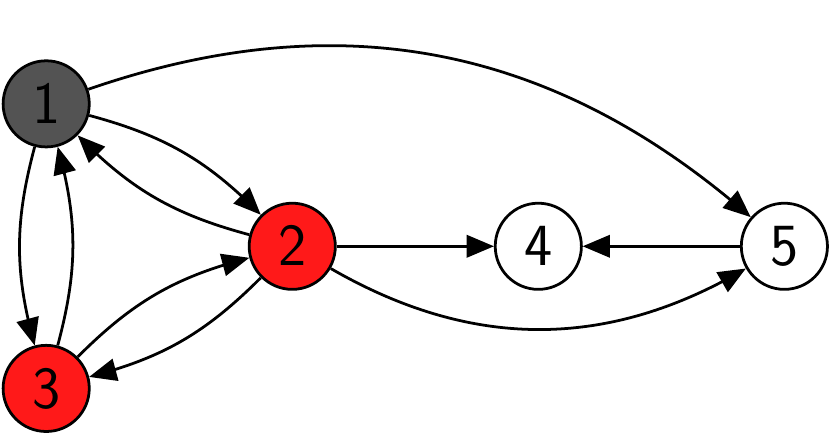} \\
    \vspace{13pt}
$\mathbf{b}^{\lambda_1}$=$(1,1,1,1,1)$\\
 $\langle b \rangle$=$1$
\end{minipage} &   
\begin{minipage}{0.3\textwidth}
  \centering
  \vspace{10pt}
  \includegraphics[scale=0.5]{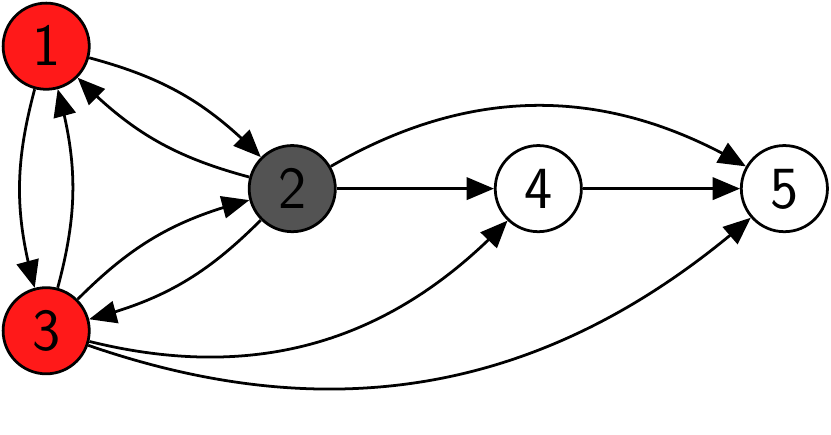} \\ \vspace{5pt}
$\mathbf{b}^{\lambda_1}$=$(1,1,1,1,1)$\\
 $\langle b \rangle$=$1$ \vspace{5pt}
\end{minipage} \\ \bottomrule[0.1em]
\multicolumn{2}{c}{\textbf{STEP} $T_{2}$}
\\ \toprule[0.05em]
\begin{minipage}{0.5\textwidth}
    \centering
    \vspace{7pt}
    \includegraphics[scale=0.5]{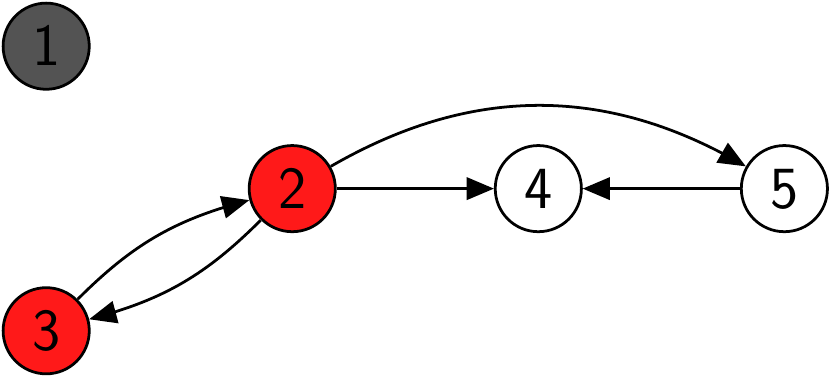} \\
    \vspace{15pt}
$\mathbf{b}^{\lambda_1}$=$(0,0.5,0.5,1,0.5)$\\
 $\langle b \rangle$=$0.5$
\end{minipage} &   
\begin{minipage}{0.3\textwidth}
  \centering
  \vspace{10pt}
  \includegraphics[scale=0.5]{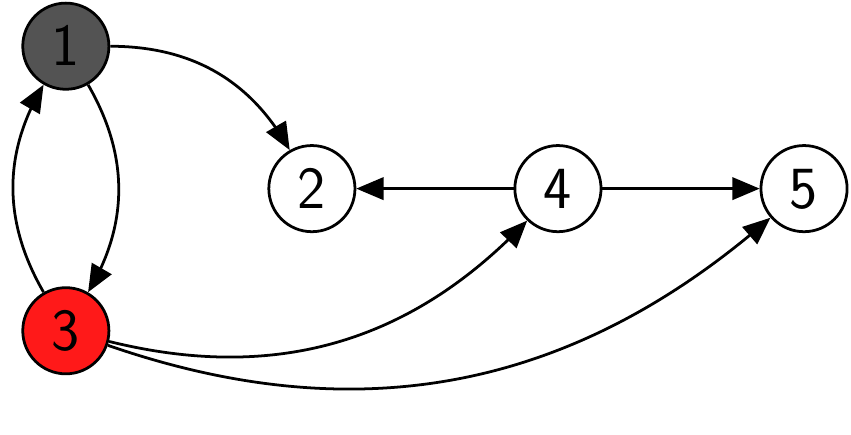} \\ \vspace{5pt}
$\mathbf{b}^{\lambda_1}$=$(0.5,1,0.5,0.5,1)$\\
 $\langle b \rangle$=$0.7$ \vspace{5pt}
\end{minipage} \\ \bottomrule[0.1em]
\multicolumn{2}{c}{\textbf{STEP} $T_{3}$}
\\ \toprule[0.05em]
\begin{minipage}{0.5\textwidth}
    \centering \vspace{7.5pt}
    \includegraphics[scale=0.5]{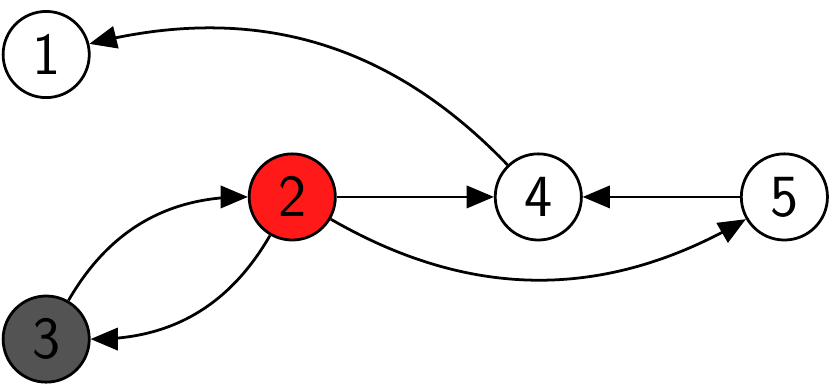} \\
    \vspace{13pt}
$\mathbf{b}^{\lambda_1}$=$(1,0.5,0.5,1,0.5)$\\
 $\langle b \rangle$=$0.7$
\end{minipage} &   
\begin{minipage}{0.3\textwidth}
  \centering
  \vspace{10pt}
  \includegraphics[scale=0.5]{figures/cost-0_2-after-first-crash.pdf} \\ \vspace{5pt}
$\mathbf{b}^{\lambda_1}$=$(0.5,1,0.5,0.5,1)$\\
 $\langle b \rangle$=$0.7$ \vspace{5pt}
\end{minipage} \\ \bottomrule[0.1em]
\multicolumn{2}{c}{\textbf{STEP} $T_{4}$}
\\ \toprule[0.05em]
\begin{minipage}{0.5\textwidth}
    \centering
    \includegraphics[scale=0.5]{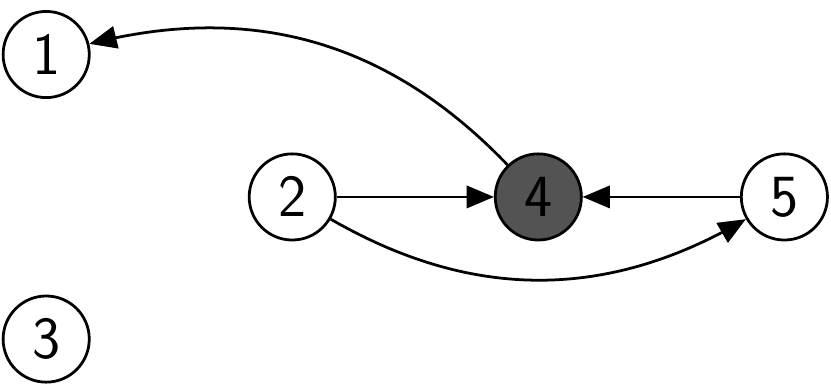} \\
    \vspace{13pt}
$\mathbf{b}^{\lambda_1}$=$(1,0,0,0,0)$\\
 $\langle b \rangle$=$0.2$ \
\end{minipage} &   
\begin{minipage}{0.3\textwidth}
  \centering
  \vspace{14pt}
  \includegraphics[scale=0.5]{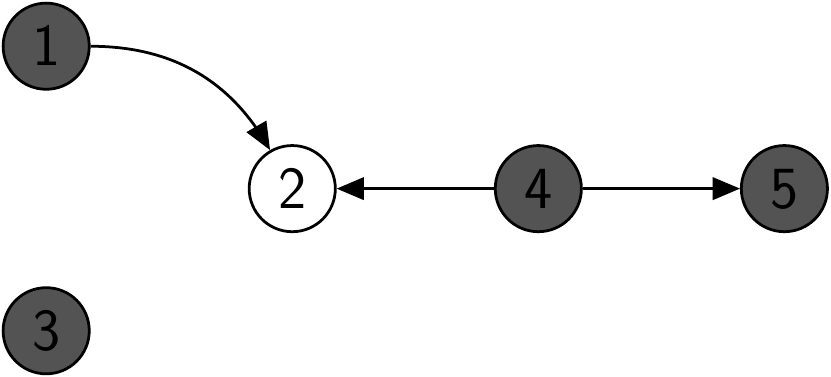} \\ \vspace{5pt}
$\mathbf{b}^{\lambda_1}$=$(0,1,0,0,0)$\\
 $\langle b \rangle$=$0.2$ 
\vspace{5pt}
\end{minipage} \\ \hline
\caption{Examples of emerging cascades of users leaving (obtained from computer simulations, see also Appendix D). The top row shows the initial configuration of a sample OSN. Subsequent rows show consecutive time steps of the evolution of the OSN, measured in network time $T_{1}$-$T_{4}$ for two different cost levels: (left column) $\tau_{1}=0$, (right column) $\tau_{2}=0.2$. (red) indicates core users, (grey) indicates users leaving at the given time step. The relative reputation $\mathbf{b}^{\lambda_{1}}$ of users and the average reputation $\mean{b}$, Eqn. (\ref{eq:satisf}), is given for each configuration.  
}
\label{tab:core-crash}
\end{longtable}
}

\section*{Appendix D: Optimal cost level}

\begin{figure}[htbp]
    \includegraphics[width=0.46\textwidth]{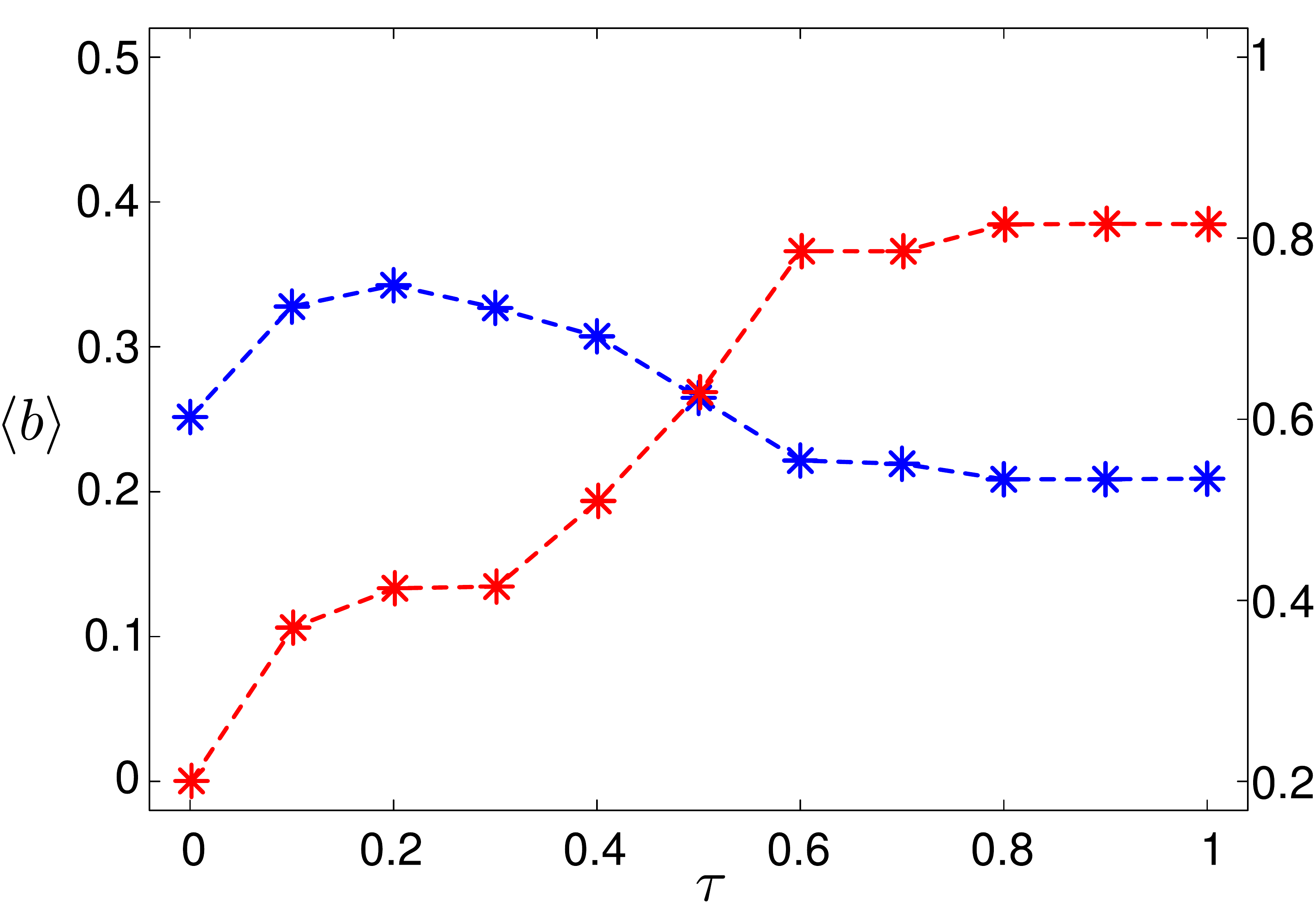} \hfill
    \includegraphics[width=0.45\textwidth]{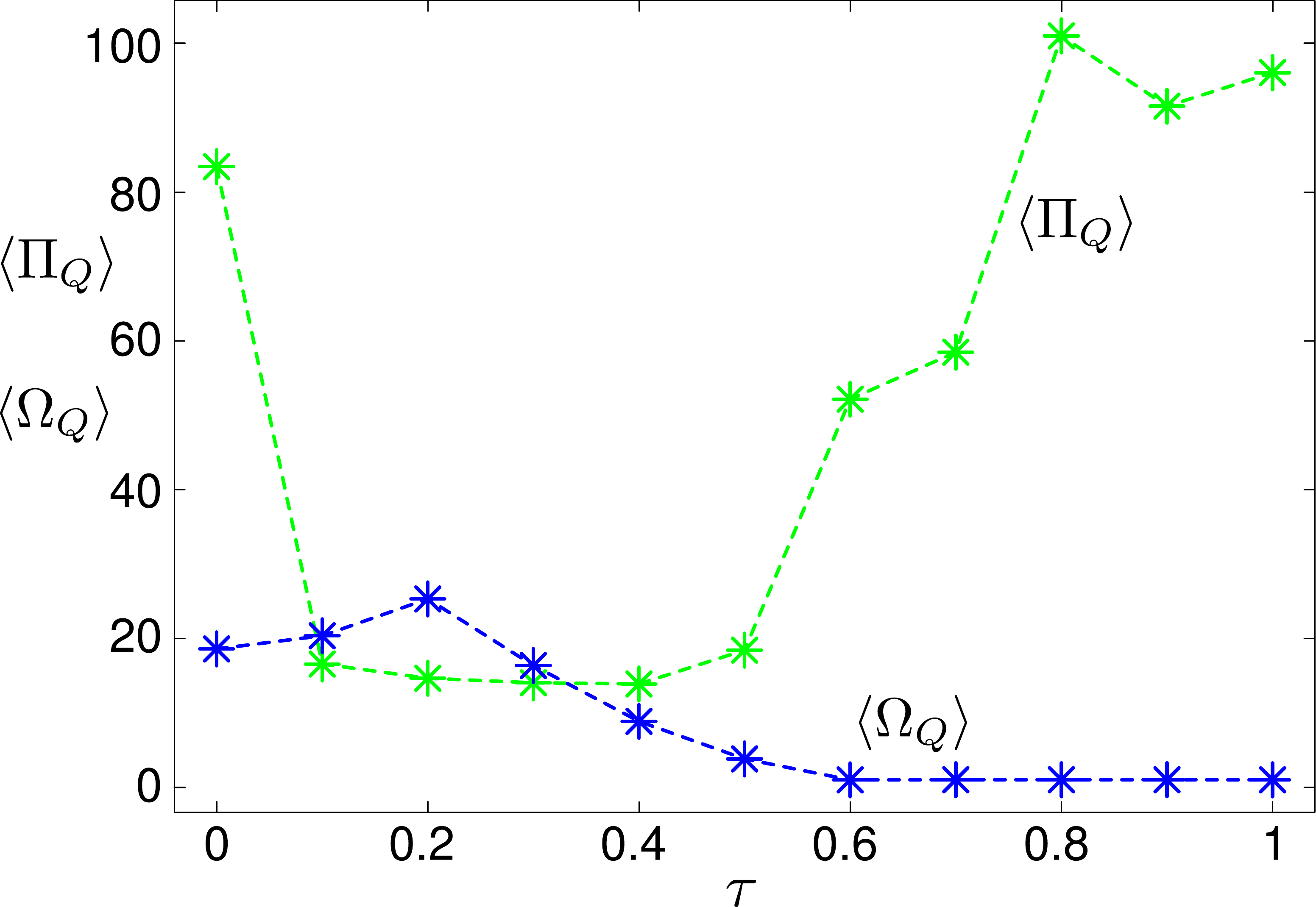}    
    \caption{(Left) Long term average benefit $\mean{b}$,
      Eqn. (\ref{eq:satisf}) (blue, left scale), and average fraction of rewired users
      over the whole simulation (red, right scale) . (Right)  Average lifetime of a
      core, $\mean{\Omega_{Q}}$ (blue), and average time to establish a
      core, $\mean{\Pi_{Q}}$ (green), both measured in network time $T$,
      as a function of the cost $\tau$. The Wilcoxon-rank test shows that
      $\mean{\Omega_{Q}}_{\tau=0.2}$ is significantly different from
      $\mean{\Omega_{Q}}_{\tau=0.1}$ (p-value = $0.02$) and from
      $\mean{\Omega_{Q}}_{\tau=0.3}$ (p-value = $10^{-8}$)  }
    \label{fig:mean-core-recovery}
\end{figure}

To illustrate how the cost, $\tau$, affects the average reputation in the
network, we simulated the dynamics of  a
small network of five nodes over $T_{\mathrm{max}}=10^{4}$ network time steps. 
The left part of Figure  \ref{fig:mean-core-recovery}, which should be compared with the left part of Figure \ref{fig:ff}, shows the 
average reputation of the OSN. We observe again that for our small
network the maximum average benefit is attained for a nonzero
cost. More precisely, we confirm that the optimal cost is $\tau_{\mathrm{max}}$=0.2, as it was for
the much larger network simulated in Figure \ref{fig:ff}, so this cannot be
attributed to  a size effect. 
Furthermore, at this optimal cost level, the network is not considerably disturbed as the average number of new users
is less than 40\% of the network size. This shows that the maximum
average benefit is not concentrated in a minority of core users, at the
expense of a large and sparsely connected periphery. 

The right part of Figure \ref{fig:mean-core-recovery}
explains why this particular value of $\tau$ appears to be the optimal one. 
Here, we have plotted two different time spans, the average lifetime of a
core, $\mean{\Omega_{Q}}$, and the average time to establish a core,
$\mean{\Pi_{Q}}$, both measured in network time $T$, as a function of the
cost $\tau$. With the exception of $\tau$=0, which only serves as a
reference case, the time $\mean{\Pi_{Q}}$ to establish a core in the OSN
monotonously increases with $\tau$ until it reaches a saturation value at
high levels of $\tau\to 1$. As explained in Sect. \ref{sec:network}, this
increase is due to the increasing number of users that leave the OSN at
each time step $T$. If $\tau$ gets too large, a considerable fraction of
the OSN is replaced, this way also destroying favorable reciprocal social
links. For large $\tau$ the average life time of the core,
$\mean{\Omega_{Q}}$, decreases for the very same reasons. For small
$\tau$, however, we observe a non-monotonous behavior, i.e. the life time
increases with $\tau$ until it reaches a maximum at $\tau_{\mathrm{max}}$=0.2. I.e. the optimal cost level balances two different dynamic effects: (a) the \emph{emergence} of a core of users that, by means of direct or indirect reciprocity, keep a relatively high reputation, and (b) the \emph{maintainence} of this core by the users. Cascades of users leaving cannot be prevented as we have explained in Appendix B. The question is how fast the OSN is able to cope with it. 
At the optimal cost level, the core not only stays alive, but also quickly recovers from the leave of core users. 
Therefore the average reputation of most of the users remain high \emph{over time}.

\section*{Appendix E: Network structure}
\label{sec:struc}

Here we present three typical snapshots of the network structure for different values of the cost $\tau$. In all three
simulations, the parameter that determines the network density is fixed to $m$=0.25 and the system size is $N$=100.  The most
immediate differences in these three snapshots are (i) the core size, which strongly decreases with increasing cost $\tau$ and
(ii) the maximum out-degree which increases with increasing cost $\tau$.  These findings are in agreement with our theoretical
discussions in Sect. \ref{sec:equilibrium} about changes in the largest eigenvalues $\lambda_{1}$ dependent on the network structure (i.e. the existence of chains and cycles).

\begin{figure}[htbp]
  \centering
  \includegraphics[width=5.3cm]{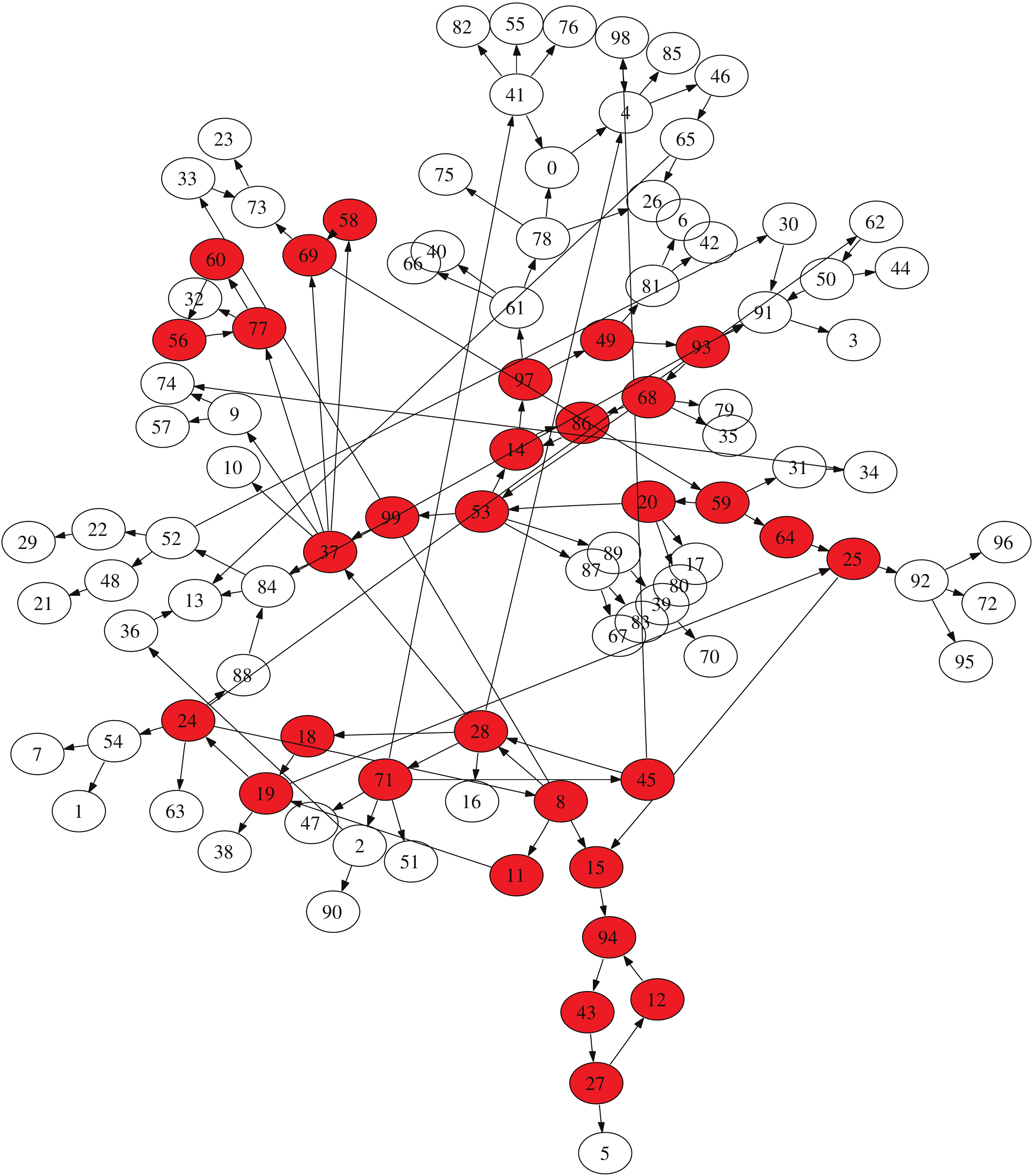}\hfill
\includegraphics[width=5.3cm]{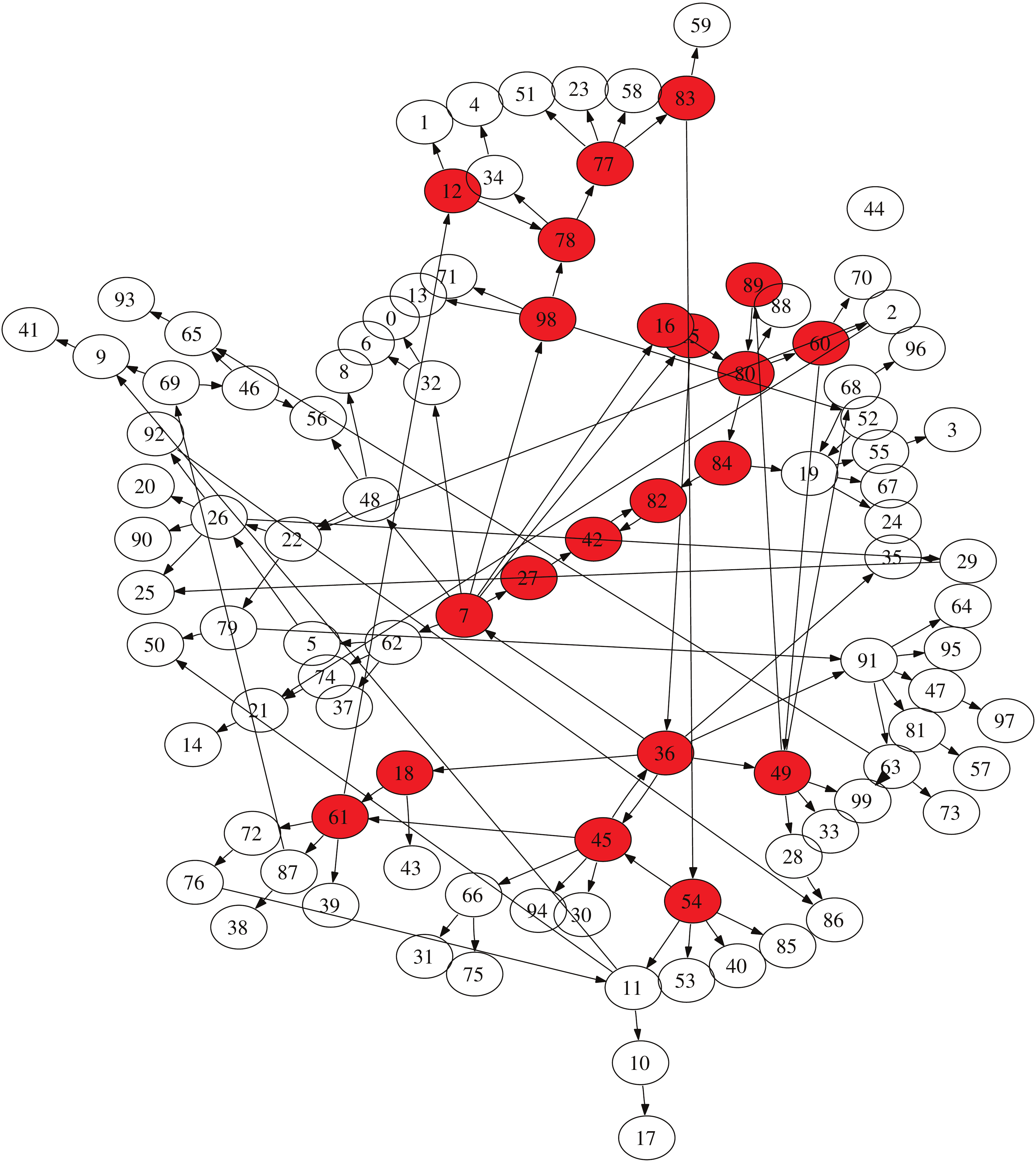}\hfill
\includegraphics[width=5.3cm]{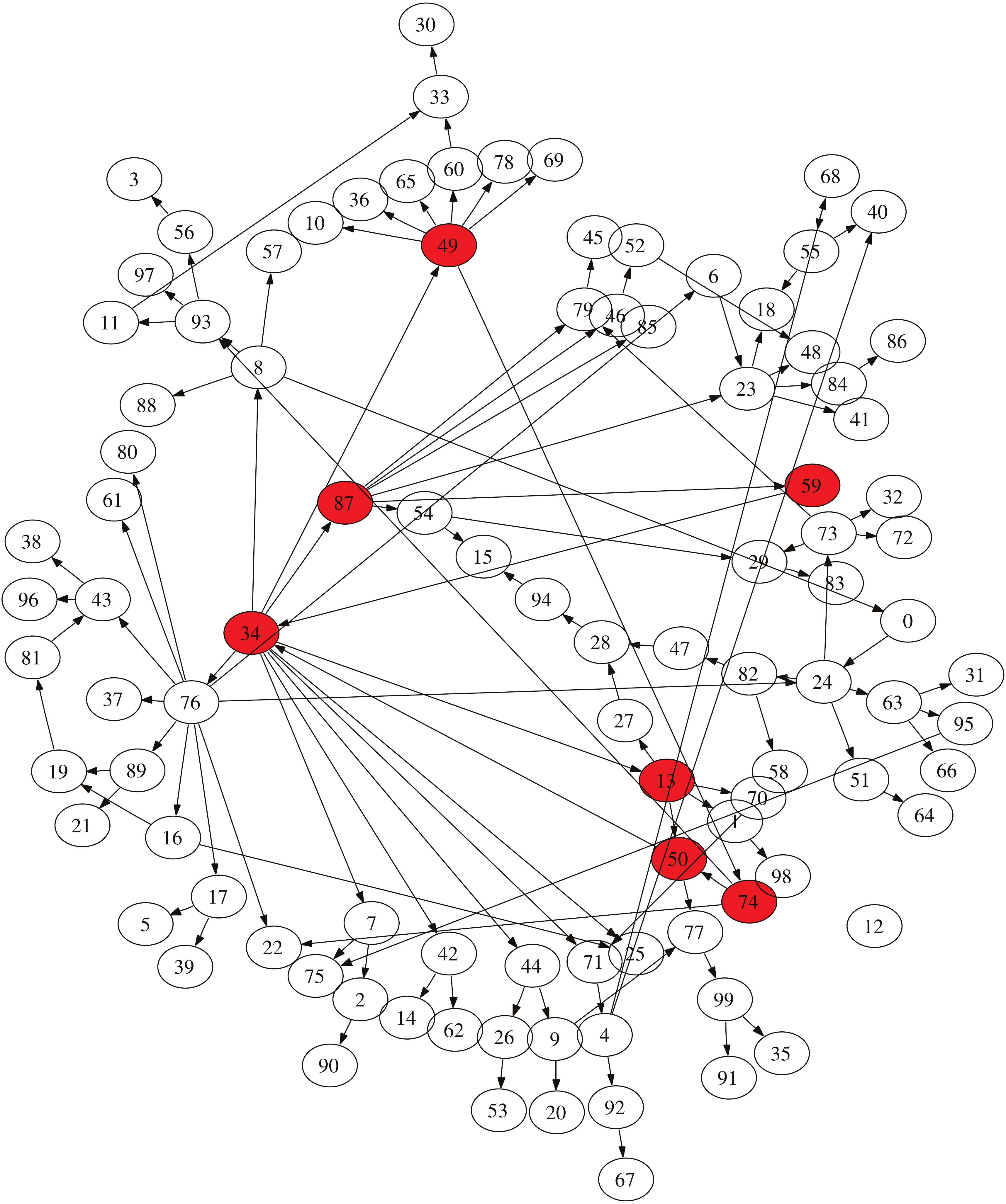}
  \caption{Sample networks for different cost $\tau$. (left) 
  $\tau$=0.0. $\lambda_1$=1.133, number of users in the core: 31.
The
largest out-degree is
$5$. In total, there are $3$ users ($71$, $28$, and $24$) with
5 outgoing links. (middle)
  $\tau$=0.2. $\lambda_1$=1.47, number of users in the core: 21
The largest outdegree is 7. User $7$
has $7$ outgoing links, user $36$ has $6$, and at least five users
($49$, $91$, $26$, $54$, and $45$) have $5$.
(right) 
  $\tau$=0.3. $\lambda_1$=$1.40$, number of users in the core: 7. 
The largest outdegree is 10. Two users ($34$ and $76$) have outdegree 10, one ($49$) has out-degree $7$  and one ($87$)
has out-degree $6$.
}
  \label{fig:networks}
\end{figure}

\end{document}